\documentclass{aa}    
\usepackage{txfonts}
\usepackage{lipsum}
\usepackage{natbib}
\usepackage{graphicx}
\usepackage{float}

\usepackage{color}

\newcommand{\Eexc}{$E_{\rm exc}$}
\newcommand{\Teff}{$T_{\rm eff}$}
\newcommand{\logg}{\rm log~g}
\newcommand{\kms}{km\,s$^{-1}$}
\newcommand{\eps}[1]{\log\varepsilon_{\rm #1}}
\newcommand{\kH}{$S_{\!\rm H}$}    
\newcommand{\eu}[5]{\mbox{${\rm #1}\,^#2{\rm #3}^{#4}_{\rm #5}$}}

\begin{document}

\title{Understanding an origin of palladium in metal-poor stars \\ based on the non-LTE analysis of \ion{Pd}{i} lines\thanks{Based on observations made with the telescopes of European Southern Observatory (ESO)}
}

\author{
L. Mashonkina\inst{1} \and 
A.~Smogorzhevskii\inst{1,2}
}

\institute{
Institute of Astronomy, Russian Academy of Sciences, RU-119017 Moscow, 
     Russia \\ \email{lima@inasan.ru}
\and M. V. Lomonosov Moscow State University, Kolmogorova st. 1, 119991, Moscow, Russia
}

\date{Received  / Accepted }

\abstract
{Palladium is one of poorly observed neutron-capture elements. Abundance determinations for stellar samples covering a broad metallicity range are needed for better understanding the mechanisms of Pd synthesis during the Galaxy evolution.  }
{We aim to obtain accurate abundances of Pd for the Sun and the sample of metal-poor stars based on the non-local thermodynamic equilibrium (non-LTE) line formation for \ion{Pd}{i}. }
{We present a new, comprehensive model atom of \ion{Pd}{i}. Abundances of Pd, Sr, Ba, and Eu were derived for 48 stars from the non-LTE analyses of high resolution and high signal-to-noise ratio spectra provided by the ESO archives. The obtained results are based on using the synthetic spectrum method with one-dimensional (1D, MARCS) model atmospheres. }
{Non-LTE leads to weakened \ion{Pd}{i} lines and positive non-LTE abundance corrections growing from 0.2~dex for the solar lines up to 0.8~dex for the lines in the most luminous star of the sample. Depending on a treatment of inelastic collisions with hydrogen atoms, the solar non-LTE abundance amounts to $\eps{\odot,Pd}$ = 1.61$\pm$0.02 to 1.70$\pm$0.02 and agrees within the error bars with the meteoritic abundance $\eps{met,Pd}$ = 1.65. Non-LTE largely removes the discrepancies in the LTE abundances between the giant and dwarf stars of similar metallicities. Palladium tightly correlates with Eu in the $-1.71 \le$ [Fe/H] $\le -0.56$ range indicating the r- and s-process contributions to Pd synthesis of approximately 70\%\ and 30\%, respectively. Palladium is of pure r-process origin in our two r-II stars, and a dominant contribution of the r-process to the Pd abundances is found for another two very metal-poor (VMP, [Fe/H] $< -2$)
stars. The two VMP stars, which are strongly enhanced with Sr relative to Ba and Eu, reveal also enhancements with Pd. We propose that the source of extra Sr and Pd in these stars are VMP, fast rotating massive stars.}
{Non-LTE is essential for 
obtaining the observational constraints to future models of the Galactic Pd evolution. }


\titlerunning{Non-LTE abundances of palladium in the Sun and metal-poor stars }
\authorrunning{Mashonkina and Smogorzhevskii}

\maketitle

\section{Introduction} \label{sec:intro}

Chemical elements beyond the Fe group are produced in neutron capture (n-capture) nuclear reactions, from the slow (s-) process and the rapid (r-) process in differing proportions for different isotopes \citep{BBFH1957}.
The s-process operates at neutron number densities of $N_n \sim 10^6 - 10^7$~cm$^{-3}$ and is represented by the main and weak s-process components \citep[see][and references therein]{Kappeler2011}. The former takes place in intermediate-mass ($0.9-8 M_\odot$) stars during their thermal pulses on the asymptotic giant branch (AGB) phase and produces heavy nuclei from gallium (Ga, $Z$ = 31) up to bismuth (Bi, $Z$ = 83).
The weak s-process component is associated with the He-burning cores of massive ($\sim 20 M_\odot$) stars and contributes to production of heavy nuclei with an atomic mass of up to $A \sim$ 90.
The r-process runs at extremely high neutron densities of $N_n > 10^{24}$~cm$^{-3}$, which can be reached in explosions of massive ($> 8 M_\odot$) stars, such as neutron stars mergers, magneto-rotational supernovae, etc. \citep[see][for a review of the astrophysical sites for the r-process]{2021RvMP...93a5002C}. The neutron capture chain of the r-process can proceed beyond Bi.

Based on a dominant contribution of either s- or r-process to their solar abundances, the chemical elements are often referred to as the s- or r-process elements. For example, Sr and Ba are the s-process elements because, according to calculations of \citet{Prantzos2020}, the s-nuclei constitute about 92\%\ and 88\%\ of the solar strontium and barium. Europium is the r-process element because the s-nuclei constitute only as little as 5\%\ of the solar europium. This paper focuses on palladium. According to \citet{Prantzos2020}, the solar palladium was produced nearly equally in the s- (45.5\%) and r-process. 
For each chemical element, a relative contribution of the r- and s-process to its synthesis varied during the Galaxy history because of different lifetimes of the sources of the r- and s-nuclei. The galactic chemical evolution models need to be verified with observations of the n-capture elements in stellar samples covering a broad metallicity range. A bulk of data is available in the literature on the best observed elements, such as Sr, Y, Zr, Ba, La, Eu, in our Galaxy \citep[for example][]{McWilliam1998, 2003A&A...397..275M, Francois2007, roe14,CERESIII} and the Galaxy satellites \citep[for example][]{2010ApJ...708..560F,  Gilmore2013, 2017A&A...608A..89M, Hill_Scl,2021AJ....162..229R}. 

Palladium is one of poorly observed n-capture elements in metal-poor (MP, [Fe/H]\footnote{In the classical notation, where [X/Y] = $\log(N_{\rm X}/N_{\rm Y})_{star} - \log(N_{\rm X}/N_{\rm Y})_{\odot}$ for each pair of elements X and Y.} $< -1$) stars. 
With a nuclear charge of $Z$ = 46 and the number of neutrons ($N_n$ = 58-60, 62, 64 for the most abundant isotopes) far from the magic numbers $N_n$ = 50, 82, 126, Pd has the lower abundance compared to, for example, abundances of Sr and Ba and is represented in stellar spectra by a very few lines of \ion{Pd}{i} located in the blue range, at $\lambda <$ 3700~\AA.
Palladium abundances were measured for the strongly r-process enhanced ([Eu/Fe] $> 1$, [Ba/Eu] $< 0$) stars \citep{Sneden2003,hill2002,2017A&A...607A..91H,HE1219} referred to as r-II stars, by definition of \citet{HERESI}. The first abundance determinations for Pd in a large sample of MP stars were reported by \citet{2012A&A...545A..31H} who concluded that the Galactic trends of elemental ratios involving Pd suggest a contribution to the Pd production from a "second/weak r-process" that differs from the classical (main) r-process with respect to neutron number densities, neutron-to-seed ratios, entropies, electron abundances. The statistics of stellar Pd measurements was increased thanks to studies of \citet{2013ApJ...768L..13P,2015A&A...579A...8W}, and \citet{2017ApJ...837....8A}.

Palladium abundance determinations made so far are all based on the assumption of local thermodynamic equilibrium (LTE). \citet{2012A&A...545A..31H} found an abundance difference between dwarfs and giants of similar metallicity and suggested that the departures from LTE could be one possible explanation. Indeed, with the ionization energy $\chi$ = 8.337~eV, palladium is ionized in the physical conditions typical for the atmospheres of MP stars and can be subject to the ultraviolet (UV) overionization caused by superthermal radiation of a non-local origin below the thresholds of the \ion{Pd}{i} low excitation levels. Non-local thermodynamic equilibrium (non-LTE = NLTE) calculations for lines of \ion{Pd}{i} were not performed to date. 

This study aims to construct the model atom of palladium and to determine, for the first time, the NLTE abundances of Pd in the Sun and the sample of 48 stars covering the $-2.92 \le$ [Fe/H] $\le -0.56$ metallicity range. The high-resolution spectra of the selected stars are available in the archives of the European Southern Observatory (ESO). We also determine the NLTE abundances of Sr, Ba, and Eu that are the best representatives of the elements produced effectively in the weak s-process, the main s-process, and the r-process, respectively. The Galactic trends of the elemental ratios among Pd, Sr, Ba, and Eu should play a key role in understanding a nucleosynthetic origin of palladium during the Galaxy evolution.

The paper is organized as follows. Section~\ref{Sect:NLTE} introduces the model atom for palladium. In Sect.~\ref{Sect:sun}, we
determine the LTE and NLTE abundances from the solar \ion{Pd}{i} lines. Section~\ref{Sect:sample} describes our stellar sample, the sources of observed spectra, and the determination of stellar atmospheric parameters. The NLTE abundances of Pd, Sr, Ba, and Eu are derived in Sect.~\ref{Sect:stars}, and the obtained Galactic abundance trends are discussed in Sect.~\ref{Sect:origin}.
Our conclusions are given in Sect.~\ref{Sect:conclusion}. 

\section{Non-LTE calculations for \ion{Pd}{i}} \label{Sect:NLTE}

\subsection{Model atom for palladium} \label{Sect:model}

\subsubsection{Energy levels}

The atomic model is built using 143 levels of neutral palladium, \ion{Pd}{i}, from the National Institute of Standards and Technology (NIST) data base\footnote{https://physics.nist.gov/asd} \citep{NIST19} and 383 levels predicted in the atomic structure calculations \citep[][files c4600e.log and c4600o.log from 2017 September 03]{Kurucz2017}, but not detected (yet) in laboratory experiments.  The high excitation levels provide close collisional coupling to the first ionization stage, \ion{Pd}{ii}, and thus play an important role in the statistical equilibrium (SE) of palladium.
The levels with low energy separation ($<$ 250~cm$^{-1} \simeq 0.03$~eV) and of the same parity were combined into superlevels. The superlevels all lie above the excitation energy \Eexc\ = 7~eV and are composed predominantly from levels predicted in calculations of the \ion{Pd}{i} atomic structure. For each superlevel, its energy was calculated from energies of individual levels with taking into account their statistical weights. The final atomic model includes 106 levels of \ion{Pd}{i}.

Singly ionized palladium is represented in the model atom by its ground state with the statistical weight $g = 10$. In fact, the lowest \ion{Pd}{ii} term, 4d$^9$~$^2$D, has two fine splitting levels, with \Eexc\ = 0 ($J = 5/2$) and \Eexc\ = 0.4388~eV ($J = 3/2$), and the partition function of \ion{Pd}{ii} varies from $U$ = 7.71 to $U$ = 10.37, when temperature grows from 5800~K to 12\,000~K. Our test calculations have revealed minor changes in the departures from LTE for lines of \ion{Pd}{i}, when $g$(\ion{Pd}{ii}) varies between 8 and 12.

\begin{figure}
 \begin{center}
  \resizebox{90mm}{!}{\includegraphics{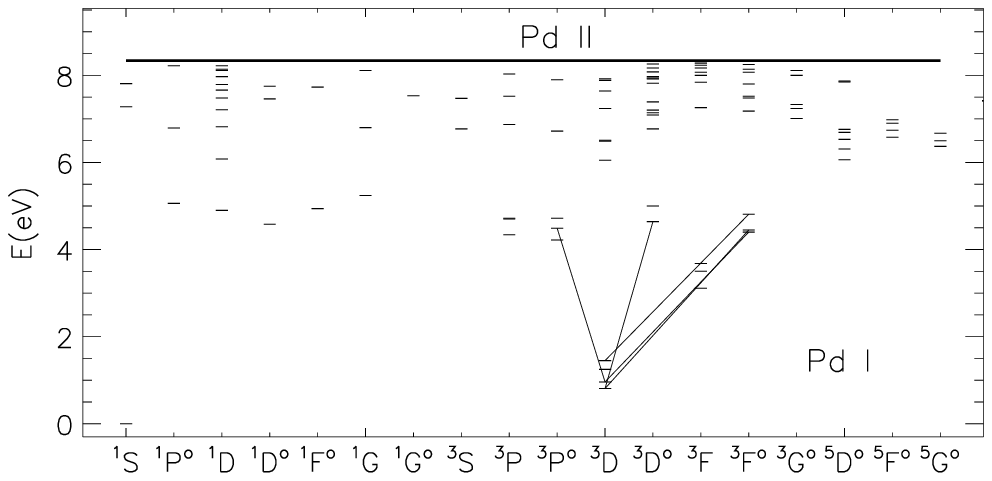}}
 \caption{Energy levels of \ion{Pd}{i}, as presented in the model atom. The spectral lines used in Pd abundance analysis arise in the transitions shown as continuous lines. The term designations correspond to atomic structure calculations of \citet{Kurucz2017}. }
 \label{fig:atom}
 \end{center}
 \end{figure}

Due to the violation of the LS-coupling, the quantum numbers of the total orbital angular momentum are not determined for many states of \ion{Pd}{i}. For example, the term of electronic configuration
 4d$^9$($^2$D$_{5/2}$)5s is indicated in the NIST data base as $^2$[5/2]. 
 In order to illustrate the \ion{Pd}{i} model atom in Fig.~\ref{fig:atom}, for all the levels, we use the term definitions as predicted by atomic structure calculations of \citet{Kurucz2017}.

 \subsubsection{Radiative transitions}
 
For 35 bound-bound (b-b) transitions, we apply oscillator strengths based on laboratory measurements of \citet{xu2006}. For 1951 b-b transitions, $gf$-values are taken from calculations of \citet[][file gf4600.low from 2017 September 06]{Kurucz2017}. The mean difference in $gf$-values between \citet{xu2006} and \citet{Kurucz2017} amounts to $-0.03\pm0.23$ for 35 common lines. 

The photoionization cross-sections are calculated in the hydrogenic approximation using the effective principal quantum number instead of the principal quantum number. In Sect.~\ref{Sect:sun}, we check an influence of variations in photoionization cross-sections for \ion{Pd}{i} on the final NLTE results.

\subsubsection{Collisional transitions}\label{Sect:coll}

Due to absence of accurate data, electron-impact excitation is computed with the semi-empirical formula of \citet{Reg1962} for the allowed transitions and adopting the effective collision strength $\Upsilon$ = 1 for the forbidden ones. When calculating electron-impact ionization rates, we rely on the \citet{1962amp..conf..375S} formula.

In the stellar parameter range with which we concern, the number density of free electrons in the atmospheres is much lower than that of neutral hydrogen atoms, with $N_{\rm e}/N_{\rm H} \sim 10^{-4}$. Therefore, the SE of any given atom should be calculated by taking into account the excitation of levels and the formation of ions by collisions with \ion{H}{i} atoms. No accurate data is available in the literature for \ion{Pd}{i} + \ion{H}{i} collisions, and we calculate the excitation and ionization rates with the \citet{Steenbock1984} formulas based on the \citet{Drawin1969} theoretical approximation. For forbidden transitions, the \ion{H}{i} collisional rates are computed using the corresponding electron-impact excitation rates, as recommended by \citet{Takeda1994}: $C_{\rm H} = C_{\rm e}\sqrt{m_{\rm e}/m_{\rm H}}N_{\rm H}/N_{\rm e}$. Here, $m_{\rm e}, m_{\rm H}$ are masses and $N_{\rm e}, N_{\rm H}$ are number densities of electrons and \ion{H}{i} atoms. Since a treatment of hydrogen collisions is very approximate, in Sects.~\ref{Sect:sun} and \ref{sect:uncertain}, we perform test calculations by introducing a scaling factor \kH\ to the Drawinian rates.

\subsection{Statistical equilibrium of \ion{Pd}{i} in stellar atmospheres of different metallicities}\label{Sect:SE}

The coupled SE and radiative transfer equations in a given atmospheric structure are solved with a modified version of the {\sc detail} code \citep{Giddings81,Butler84}. The revised opacity package was described by \citet{mash_fe}. Everywhere in this study, we use the plane-parallel (1D) model atmospheres obtained by the interpolation in the MARCS\footnote{http://marcs.astro.uu.se} grid \citep{Gustafssonetal:2008} for given effective temperature (\Teff), surface gravity (\logg), and iron abundance [Fe/H].

\begin{figure} 
	\hspace{-3mm}
	\centering
  \resizebox{88mm}{!}{\includegraphics{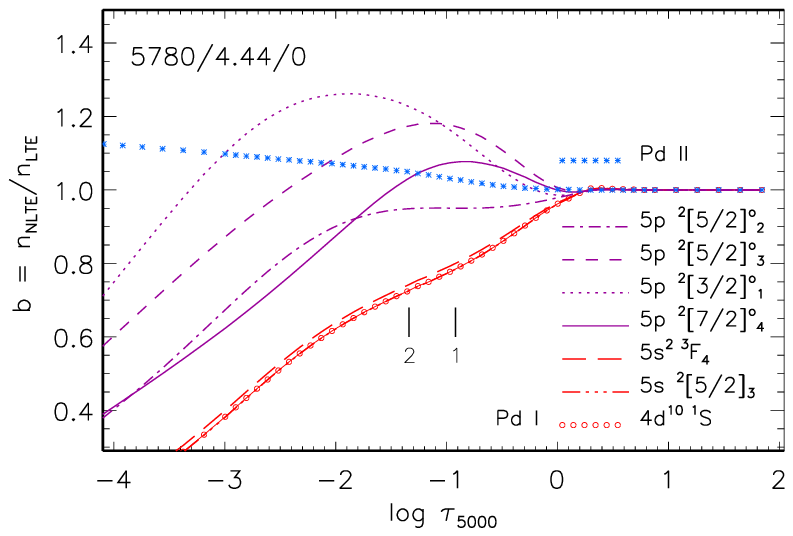}}
	\caption{Departure coefficients, b, for the selected levels of \ion{Pd}{i} as a function of $\log\tau_{5000}$ in the model atmosphere 5780/4.44/0.
The tick marks indicate the location of line center optical depth unity for the  \ion{Pd}{i} 3242 (1) and 3404~\AA\ (2) lines.}
	\label{fig:bf}
\end{figure}

Figure~\ref{fig:bf} displays the departure coefficients, ${\rm b = n_{NLTE}/n_{LTE}}$, of the selected atomic levels in the model atmosphere with \Teff\ = 5780~K, \logg\ = 4.44, [Fe/H] = 0 (5780/4.44/0).
Here, ${\rm n_{NLTE}}$ and ${\rm n_{LTE}}$ are the statistical equilibrium and thermal (Saha-Boltzmann) number densities, respectively. Everywhere in the model, palladium is strongly ionized, with $N$(\ion{Pd}{i})/$N$(Pd) $< 0.05$.
Number density of the minority species is sensitive to deviations of the intensity of the ionizing radiation from the Planck function. 
Superthermal radiation in the 3000--4000~\AA\ spectral range, where the ionization thresholds of the \ion{Pd}{i}
\eu{4d^9(^2D_{5/2})5p}{2}{[5/2]}{\circ}{}, \eu{4d^9(^2D_{3/2})5p}{2}{[1/2]}{\circ}{}, \eu{4d^9(^2D_{3/2})5p}{2}{[5/2]}{\circ}{}  levels are located, leads to their overionization and the effect is transferred to other levels via the b-b transitions. On the other side, radiative pumping from the low-excitation even levels tends to overpopulate the odd levels \eu{4d^95p}{2}{[7/2]}{\circ}{}, $^2$[3/2]$^\circ$, $^2$[5/2]$^\circ$ ($^3$F$^\circ$, $^3$P$^\circ$, $^3$D$^\circ$ in Fig.~\ref{fig:atom}). 
The net effect is depleted populations (b $< 1$) of the \Eexc\ $<$ 4~eV levels and enhanced populations (b $> 1$) of the 4~eV $<$ \Eexc\ $<$ 5~eV levels in the line formation layers.

Similar behavior of the departure coefficients was found in the MP models. However, the intensity of the UV ionizing radiation is greater there due to the lower metal opacity, while effect of radiative pumping is weaker due to the lower Pd abundance. As a result, all the \ion{Pd}{i} levels have depleted populations in the line formation layers.

For all stellar atmosphere parameters, with which we concern in this study, NLTE leads to weakened \ion{Pd}{i} lines because, for each line, the lower level of the corresponding transition has depleted population (b$_{\rm low} < 1$) in the line formation layers and the line source function exceeds the Planck function due to b$_{\rm low} <$ b$_{\rm up}$.

\begin{table*}
	\centering
	\caption{Solar LTE and NLTE (two scenarios: with \kH\ = 0.1 and 1) abundances from individual lines of \ion{Pd}{i}.}
	\label{tab:sun}
	\begin{tabular}{lcrcccc}
		\hline \noalign{\smallskip}
	\multicolumn{1}{c}{$\lambda$} & \Eexc & lg $gf$ & Transition  & \multicolumn{3}{c}{$\eps{}$}  \\
		\cline{5-7}
	\multicolumn{1}{c}{(\AA)}     &  (eV) &         &          & LTE & \kH\ = 0.1 & \kH\ = 1 \\
		\noalign{\smallskip} \hline \noalign{\smallskip}
3242.70 &  0.814 & 0.07 & \eu{4d^95s}{2}{[5/2]}{}{3} -- \eu{4d^95p}{2}{[5/2]}{\circ}{3} & 1.51 & 1.73 & 1.64 \\
3404.58 &  0.814 & 0.33 & \eu{4d^95s}{2}{[5/2]}{}{3} -- \eu{4d^95p}{2}{[7/2]}{\circ}{4} & 1.49 & 1.70 & 1.61  \\
3609.55 &  0.962 & 0.13 & \eu{4d^95s}{2}{[5/2]}{}{2} -- \eu{4d^95p}{2}{[7/2]}{\circ}{3} & 1.47 & 1.68 & 1.59  \\
\cline{5-7}   \noalign{\smallskip}
	&        &      &                   \multicolumn{1}{r}{Mean}                    & 1.49 & 1.70 & 1.61 \\
    &        &      &                  \multicolumn{1}{r}{$\sigma_{\rm Pd}$}                 & 0.02 & 0.02 & 0.02 \\
\noalign{\smallskip}\hline \noalign{\smallskip}
3516.94 & 0.962 & --0.21 & \eu{4d^95s}{2}{[5/2]}{}{2} -- \eu{4d^95p}{2}{[3/2]}{\circ}{1} & 1.52$^1$ & 1.72$^1$ & 1.63$^1$ \\
3690.34 & 1.453 & --0.58 & \eu{4d^95s}{2}{[3/2]}{}{2} -- \eu{4d^95p}{2}{[5/2]}{\circ}{2} & 1.64 & 1.79 & 1.70 \\
\noalign{\smallskip}\hline 
\end{tabular}

$^1$ using the local continuum level.
\end{table*}

\section{Solar abundance of Pd} \label{Sect:sun}

\subsection{Used methods}

We analyze five best observed and least blended lines of \ion{Pd}{i} in the solar disc-center intensity spectrum by \citet{Delbouille1973}. They are listed in Table~\ref{tab:sun}. 
The Pd abundances were determined using the synthetic spectrum method, that is, by automatically fitting the theoretical spectrum to the observed one. The synthetic intensity spectrum was calculated with the code \textsc{siu} \citep{Reetz}. Using the departure coefficients from the code {\sc detail}, \textsc{siu} calculates lines of \ion{Pd}{i} taking into account the NLTE effects, while lines of other elements are computed under the LTE assumption.
The line atomic parameters for calculating the synthetic spectrum were taken from the Vienna Atomic Line Database \citep[VALD,][]{2015PhyS...90e4005R}. For the \ion{Pd}{i} lines, we adopted $gf$-values from  \citet{xu2006}. Palladium is represented in the nature by six isotopes. The fraction of the only isotope with the odd atomic mass, $^{105}$Pd, constitutes 22.33\%\ in the solar system matter \citep{Lodders2021}. According to \citet{Engleman1998}, isotopic shifts and hyperfine splitting (HFS) the \ion{Pd}{i} levels are minor, and they are not taken into account in our calculations.

As in our previous studies, we use the canonical solar parameters: \Teff\ = 5780~K, \logg\ = 4.44, microturbulent velocity $\xi_{t}$ = 0.9~\kms\ and the classical atmospheric model from the MARCS database.

\begin{figure*}  
	\centering
  \resizebox{150mm}{!}{\includegraphics{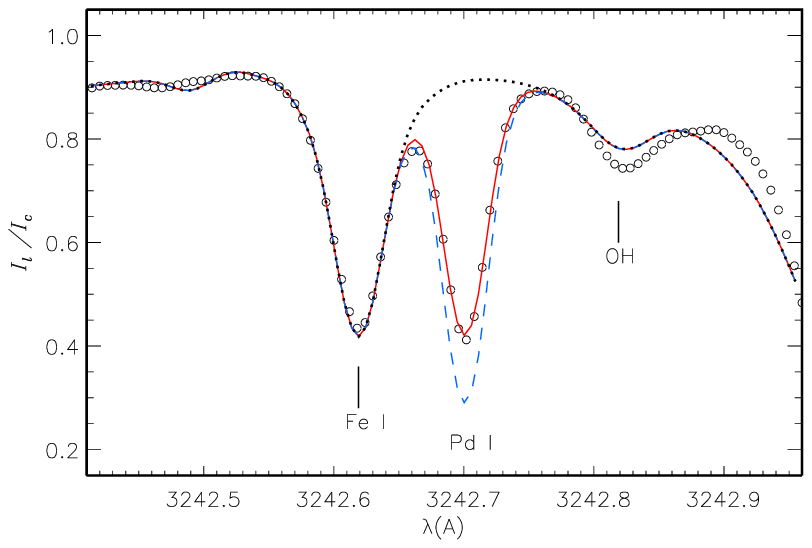} \includegraphics{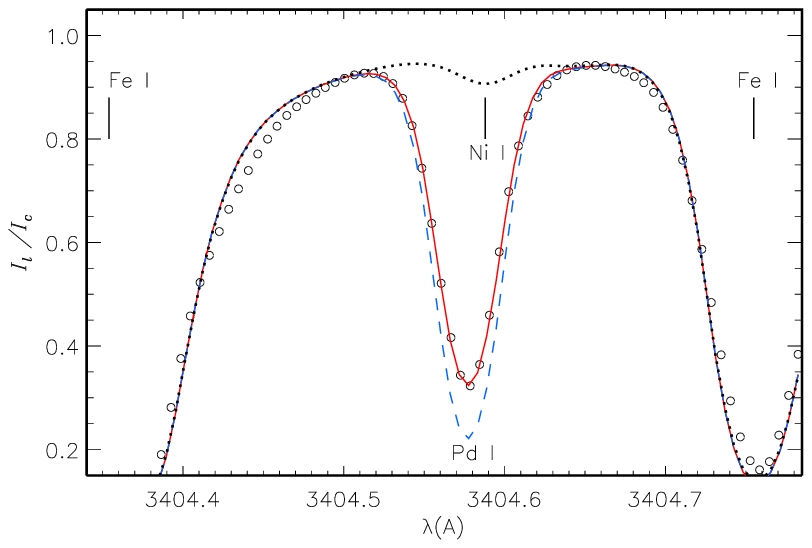}} \resizebox{150mm}{!}{\includegraphics{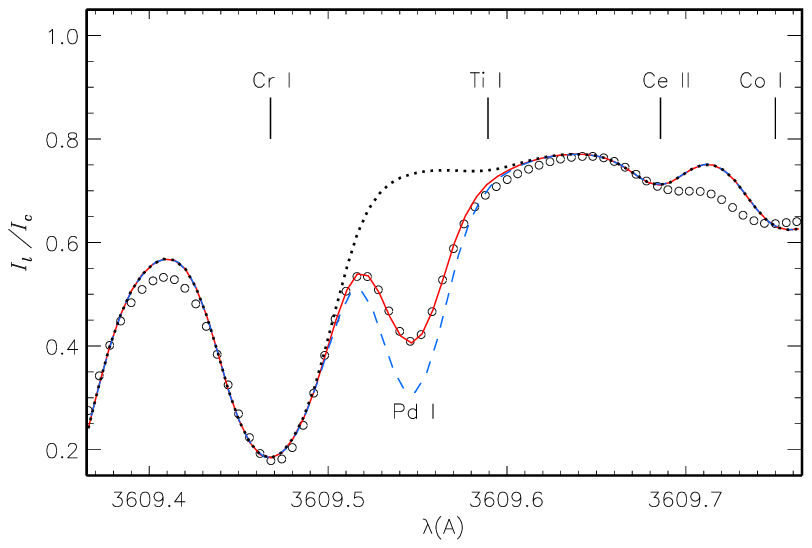} \includegraphics{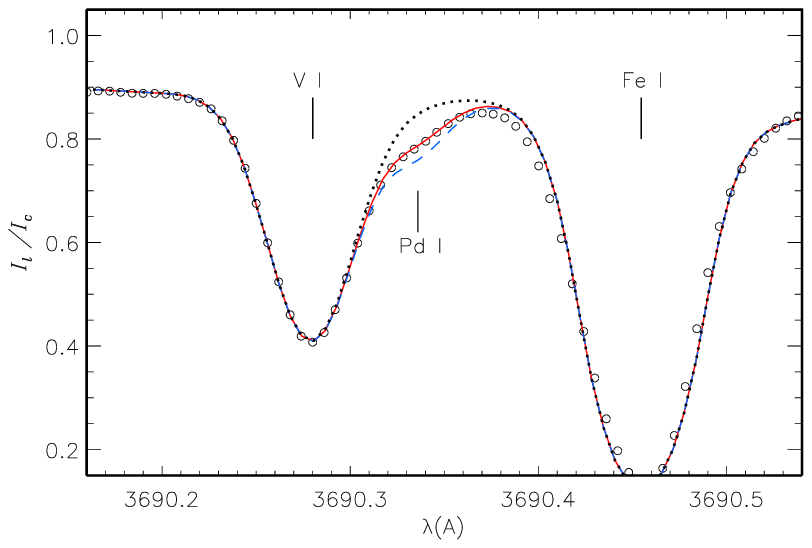}}
	\caption{Best NLTE (\kH\ = 0.1) fits (continuous curves) of the \ion{Pd}{i} lines in the solar disk-center intensity spectrum \citep[][open circles]{Delbouille1973}. The obtained NLTE abundances are indicated in Table~\ref{tab:sun}. In each panel, the dashed curve shows the LTE profile computed with the NLTE abundance derived from this line and the dotted curve is the synthetic spectrum without the contribution of the \ion{Pd}{i} line.
	}
	\label{fig:pd1}
\end{figure*}

\begin{figure}  
	\centering
  \resizebox{88mm}{!}{\includegraphics{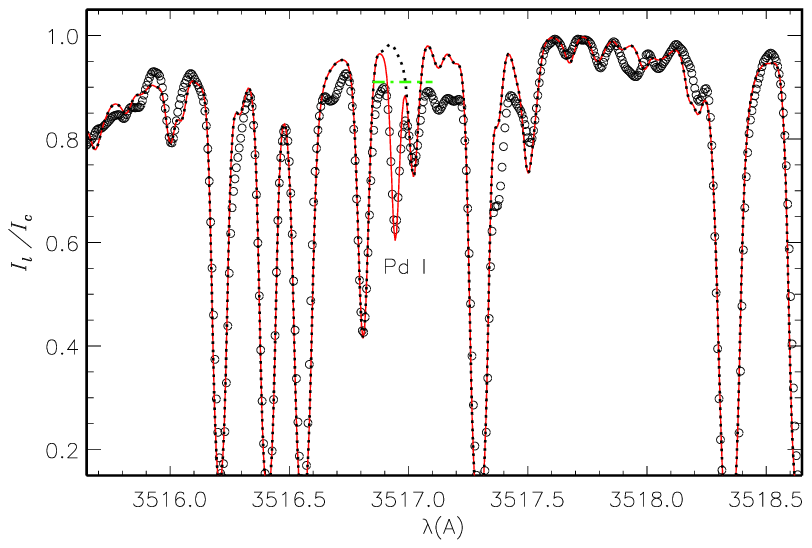}}
	\caption{Best fit (continuous curve) of the 3515.8 -- 3518.5~\AA\ range in the solar disk-center intensity spectrum \citep[][open circles]{Delbouille1973}. The dotted curve is the synthetic spectrum without the contribution of the \ion{Pd}{i} 3516~\AA\ line. The dash-dotted line shows a local continuum level that was adopted for \ion{Pd}{i} 3516~\AA\ when deriving the element abundance listed in Table~\ref{tab:sun}.
	}
	\label{fig:pd3516}
\end{figure}

\subsection{Abundances from individual \ion{Pd}{i} lines}

The Pd abundances were derived under the LTE assumption and in the two NLTE line formation scenarios with different treatment of collisions with \ion{H}{i} atoms, namely, by applying \kH\ = 0.1 or 1 to the Drawinian rates. The obtained results are presented in Table~\ref{tab:sun}. Here, we describe particular fitting procedures for the \ion{Pd}{i} lines.

The \ion{Pd}{i} 3404.579~\AA\ line is the most reliable abundance indicator. It is located in the overlapping wings of two strong blends, namely, \ion{Fe}{i} 3404.270~\AA\ + \ion{Fe}{i} 3404.301~\AA\ + \ion{Fe}{i} 3404.354~\AA\ and \ion{Fe}{i} 3404.755~\AA\ + \ion{Zr}{ii} 3404.827~\AA\ (Fig.~\ref{fig:pd1}). To begin, we fit the first of these blends by allowing the Fe abundance and the macroturbulent velocity $V_{\rm mac}$ to vary. With fixed Fe abundance,  the 3404.1--3405.2~\AA\ spectral range was fitted treating the Pd abundance and $V_{\rm mac}$ as free parameters. The \ion{Ni}{i} 3404.588~\AA\ line has a minor impact on the derived Pd abundance. 

The \ion{Pd}{i} 3242.700~\AA\ line is overlapping with \ion{Fe}{i} 3242.619~\AA\ and close to the molecular OH 3242.819~\AA\ line (Fig.~\ref{fig:pd1}). Abundances of Fe and O were adjusted in advance in order to reproduce the absorption in the corresponding lines. 

The \ion{Pd}{i} 3609.547~\AA\ line is overlapping with \ion{Fe}{i} 3609.462~\AA\ and \ion{Cr}{i} 3609.468~\AA\  (Fig.~\ref{fig:pd1}), and all the three lines lie in the wing of very strong \ion{Fe}{i} 3608.859~\AA\ line. In order to fit the wing of \ion{Fe}{i} 3608.859~\AA\ around 3609.65~\AA, the Fe abundance was reduced by 0.07~dex compared to the Solar System value $\eps{Fe,met}$ = 7.45 \citep{Lodders2021}. We use the abundance scale where $\eps{H}$ = 12. 

The \ion{Pd}{i} 3690.336~\AA\ line is blended with much stronger \ion{V}{i} 3690.29~\AA\ line (Fig.~\ref{fig:pd1}). The blend was fitted by taking into account the HFS components of the \ion{V}{i} line and reducing the V abundance by 0.23~dex compared to $\eps{V,met}$ = 3.95 \citep{Lodders2021}. 

An absorption of unknown origin in the 3516.9--3517.2~\AA\ range, where the \ion{Pd}{i} 3516.944~\AA\ line is located (Fig.~\ref{fig:pd3516}), makes uncertain an abundance determination from this line. Using the local continuum level shown in Fig.~\ref{fig:pd3516} by the dash-dotted line, we obtain a 0.16~dex lower Pd abundance compared to the case, where the continuum level is determined for an extended region from 3516.0 to 3518.5~\AA. 

The mean abundances and their errors $\sigma_{\rm Pd}$ were calculated without including \ion{Pd}{i} 3516 and 3690~\AA. Here, $\sigma = \sqrt{\Sigma(\overline{x}-x_i)^2 / (N-1)}$ is the dispersion in the single line measurements around the mean and $N$ is the number of lines measured.
 In both NLTE scenarios, $\eps{Pd,\odot}$(\kH\ = 0.1) = 1.70$\pm$0.02 and $\eps{Pd,\odot}$(\kH\ = 1) = 1.61$\pm$0.02 are consistent within the error bars with the meteoritic value $\eps{Pd,met}$ = 1.65$\pm$0.02 \citep{Lodders2021}, while the LTE abundance is lower by 0.16~dex.

\subsection{Uncertainties in derived NLTE abundances}

\begin{table}
	\centering
	\caption{Error estimates for NLTE calculations of \ion{Pd}{i} 3404~\AA\ in the Sun: changes in $\eps{}$, when varying atomic data in the model atom compared to the standard recipe (Sect.~\ref{Sect:model}, \kH\ = 0.1 for all b-b transitions). }
	\label{tab:sun_errors}
	\begin{tabular}{lc}
		\hline \noalign{\smallskip}
Variations in the model atom &  $\eps{Pd}$ -- 1.70 \\
		\noalign{\smallskip} \hline \noalign{\smallskip}
Collisions with \ion{H}{i} atoms: & \\
 \ \ \kH\ = 0 &   +0.04 \\
 \ \ \kH\ = 0 for forbidden transitions & +0.04 \\
 \ \ \kH\ = 1 & --0.09 \\
Collisions with electrons: & \\
 \ \ rate coefficients $\times$ 10 & --0.04 \\
 \ \ rate coefficients $\times$ 0.1 & 0.00 \\
Photoionizations: & \\
 \ \ cross-sections $\times$ 10 & +0.01 \\
 \ \ cross-sections $\times$ 0.1 & 0.00 \\
 LTE & --0.21 \\
\noalign{\smallskip}\hline 
\end{tabular}
\end{table}

As can be seen from Table~\ref{tab:sun}, the departures from LTE are similar for the investigated lines, and we choose \ion{Pd}{i} 3404~\AA\ in order to check the sensitivity of NLTE abundances to variations in atomic model. The results of our tests are summarized in Table~\ref{tab:sun_errors}. As expected, increasing collisional rates reduces the NLTE effects.
The abundance shift is larger for collisions with \ion{H}{i} atoms ($-0.09$~dex) than  electrons ($-0.04$~dex). In the presence of collisions with \ion{H}{i} atoms with \kH\ = 0.1, reducing electron-impact excitation rates by a factor of 10 produces no effect on the NLTE results.

It seems surprising that decreasing or increasing photoionization cross-sections by a factor of 10 has only a little effect on the derived NLTE abundance. This can be explained as follows. 
Close collisional coupling of the \ion{Pd}{i} high-excitation levels to the ground state of \ion{Pd}{ii} and the combined photon losses in many b-b transitions to the lower excitation levels siphon an efficient flow of electrons downward and 
compensate efficiently the population loss caused by increasing the photoionization rates.

\subsection{Comparison with other studies}

All determinations of the solar Pd abundance available in the literature were made under the LTE assumption. Therefore,  our LTE value, $\eps{Pd,\odot}$(LTE) = 1.49$\pm$0.02 is used for comparisons. 
\citet{1982A&A...108..127B} applied $gf$-values based on their own lifetime measurements and determined $\eps{Pd}$ = 1.68$\pm$0.04 from eight lines. Using updated $gf$-values and five lines of \ion{Pd}{i} with equivalent widths (EWs) measured by \citet{1982A&A...108..127B}, 
\citet{xu2006} derived $\eps{Pd}$ = 1.66$\pm$0.04. Both studies relied on the semi-empirical solar model atmosphere of \citet[][HM]{1974SoPh...39...19H} that is hotter in the line formation layers compared to the theoretical MARCS model atmosphere.
 This explains why they obtained the higher Pd abundance compared to our LTE value. Our result is fully consistent with $\eps{Pd}$ = 1.49 derived by \citet{Grevesse2015} using two lines of \ion{Pd}{i} and the MARCS model atmosphere. The same paper reports the higher Pd abundance, by 0.06~dex, from calculations with the three-dimensional (3D) model atmosphere. With slightly revised EWs and the same 3D model atmosphere, \citet{Asplund2021} obtained $\eps{Pd}$ = 1.57$\pm$0.10.
 
 The 3D-LTE abundances of solar palladium are higher than our 1D-LTE value, by 0.08~dex, however, they are lower than the meteoritic Pd abundance, by 0.08~dex, indicating the need for NLTE modeling the \ion{Pd}{i} lines. 
 Applying the treated NLTE method, we achieved an agreement of the solar photosphere and meteoritic abundances, although the uncertainty remains due to poorly known \ion{Pd}{i} + \ion{H}{i} collisions.
 
\begin{table*}  
	\centering
\scriptsize{	
	\caption{\label{Tab:abundances1} Atmospheric parameters for the sample stars and palladium NLTE and LTE abundances, with the [Pd/Fe] ratios computed using $\eps{Pd,met}$ = 1.65 \citep{Lodders2021}. }
	\begin{tabular}{lccccllccr}
		\hline \noalign{\smallskip}
		\multicolumn{1}{c}{Star} & \Teff & \logg & [Fe/H] & $\xi_t^1$ & \multicolumn{2}{c}{$\eps{Pd}$} & NLTE -- LTE & \multicolumn{2}{c}{[Pd/Fe]}  \\
		\cline{6-7}
		\cline{9-10} \noalign{\smallskip}
		&  (K)  &      &       &    & \multicolumn{1}{c}{NLTE} & \multicolumn{1}{c}{LTE} & (dex) & \multicolumn{1}{c}{NLTE} & \multicolumn{1}{c}{LTE} \\
		\noalign{\smallskip}  \hline \noalign{\smallskip}
		\multicolumn{10}{c}{\Teff (IRFM) from \citet{Casagrande2010,Casagrande2011,2009A&A...497..497G, 1999A&AS..139..335A} } \\
		BD -01 2582 & 5220 & 2.82 & --2.20 & 1.6 & ~~0.13(0.07) & --0.41(0.07) &  0.54 &  0.68 & ~~0.14  \\
		CD -33 3337 & 6000 & 3.88 & --1.34 & 1.4 & ~~0.68       & ~~0.24       &  0.44 &  0.37 & --0.07  \\
		CD -57 1633 & 6050 & 4.36 & --0.88 & 1.2 & ~~1.13(0.05) & ~~0.79(0.04) &  0.34 &  0.36 & ~~0.02  \\
		HD 3567     & 6180 & 4.06 & --1.14 & 1.4 & ~~1.10(0.10) & ~~0.68(0.09) &  0.42 &  0.59 & ~~0.17  \\
		HD 22879    & 5940 & 4.28 & --0.89 & 1.2 & ~~1.32(0.07) & ~~0.97(0.09) &  0.35 &  0.56 & ~~0.21  \\
		HD 25704    & 5940 & 4.23 & --0.88 & 1.2 & ~~1.14(0.08) & ~~0.78(0.07) &  0.36 &  0.37 & ~~0.01  \\
		HD 63077    & 5870 & 4.19 & --0.82 & 1.2 & ~~1.33(0.08) & ~~0.96(0.09) &  0.37 &  0.50 & ~~0.13  \\
		HD 63598    & 5900 & 4.21 & --0.87 & 1.2 & ~~1.30(0.05) & ~~0.93(0.07) &  0.37 &  0.52 & ~~0.15  \\
		HD 76932    & 6000 & 4.14 & --0.88 & 1.3 & ~~1.36(0.06) & ~~0.99(0.07) &  0.37 &  0.59 & ~~0.22  \\
		HD 83212    & 4550 & 1.00 & --1.44 & 1.9 & ~~0.54       & --0.03       &  0.57 &  0.33 & --0.24  \\   
		HD 105004   & 5880 & 4.40 & --0.79 & 1.1 & ~~1.27(0.06) & ~~0.94(0.07) &  0.33 &  0.41 & ~~0.08  \\
		HD 106038   & 6115 & 4.29 & --1.34 & 1.3 & ~~1.03(0.09) & ~~0.66(0.06) &  0.37 &  0.72 & ~~0.35  \\
		HD 111980   & 5910 & 4.00 & --1.09 & 1.3 & ~~1.22(0.06) & ~~0.82(0.08) &  0.40 &  0.66 & ~~0.26  \\
		HD 113679   & 5680 & 4.02 & --0.64 & 1.2 & ~~1.38(0.11) & ~~1.05(0.12) &  0.33 &  0.37 & ~~0.04  \\
		HD 120559   & 5450 & 4.43 & --0.98 & 0.9 & ~~1.17(0.09) & ~~0.82(0.11) &  0.35 &  0.50 & ~~0.15  \\
		HD 121004   & 5910 & 4.61 & --0.66 & 1.1 & ~~1.59(0.07) & ~~1.34(0.12) &  0.25 &  0.60 & ~~0.35  \\
		HD 126587   & 4870 & 2.21 & --2.74 & 1.7 & --0.80(0.22) & --1.43(0.21) &  0.63 &  0.29 & --0.34  \\
		HD 126681   & 5640 & 4.65 & --1.13 & 0.9 & ~~1.18(0.07) & ~~0.86(0.06) &  0.32 &  0.66 & ~~0.34  \\
		HD 132475   & 5810 & 3.86 & --1.41 & 1.3 & ~~0.84(0.09) & ~~0.41(0.11) &  0.43 &  0.60 & ~~0.17  \\
		HD 160617   & 6050 & 3.90 & --1.65 & 1.4 & ~~0.72       & ~~0.25       &  0.47 &  0.72 & ~~0.25  \\
		HD 165195   & 4420 & 0.97 & --2.15 & 2.0 & ~~0.08(0.06) & --0.72(0.11) &  0.80 &  0.58 & --0.22  \\
		HD 166913   & 6270 & 4.27 & --1.43 & 1.4 & ~~0.64       & ~~0.28       &  0.36 &  0.42 & ~~0.06  \\
		HD 175179   & 5850 & 4.30 & --0.71 & 1.2 & ~~1.53(0.09) & ~~1.19(0.11) &  0.34 &  0.59 & ~~0.25  \\
		HD 186478   & 4700 & 1.70 & --2.30 & 1.8 & ~~0.04       & --0.68       &  0.72 &  0.69 & --0.03  \\
		HD 188510   & 5560 & 4.68 & --1.51 & 0.8 & ~~0.58(0.09) & ~~0.30(0.09) &  0.28 &  0.44 & ~~0.16  \\
		HD 189558   & 5765 & 3.87 & --1.12 & 1.3 & ~~1.23(0.06) & ~~0.83(0.06) &  0.40 &  0.70 & ~~0.30  \\
		HD 195633   & 6130 & 3.92 & --0.56 & 1.5 & ~~1.39(0.07) & ~~1.02(0.07) &  0.37 &  0.30 & --0.07  \\
		HD 204543   & 4670 & 1.41 & --1.59 & 1.9 & ~~0.21(0.12) & --0.40(0.16) &  0.61 &  0.15 & --0.46  \\
		\multicolumn{10}{c}{\Teff\ from VJHK photometry (this study)} \\
		BD +20 0571 & 6130 & 4.05 & --0.91 & 1.4 & ~~1.59(0.06) & ~~1.21(0.08) &  0.38 &  0.85 & ~~0.47  \\
		CD -45 3283 & 6220 & 4.57 & --0.95 & 1.2 & ~~1.91(0.06) & ~~1.58(0.07) &  0.33 &  1.21 & ~~0.88  \\
		HD 60319    & 6170 & 4.09 & --0.86 & 1.4 & ~~1.35(0.06) & ~~0.96(0.05) &  0.39 &  0.56 & ~~0.17  \\
		HD 103723   & 6180 & 4.18 & --0.81 & 1.4 & ~~1.41(0.04) & ~~1.05(0.05) &  0.36 &  0.57 & ~~0.21  \\
		HD 122956   & 4720 & 1.53 & --1.67 & 1.8 & ~~0.50(0.13) & --0.04(0.15) &  0.54 &  0.52 & --0.02  \\
		HD 205650   & 6080 & 4.40 & --1.17 & 1.2 & ~~1.24(0.06) & ~~0.87(0.06) &  0.37 &  0.76 & ~~0.39  \\
		\multicolumn{10}{c}{Spectroscopic \Teff\ from \citet{2003A&A...397..275M}} \\
		HD 29907    & 5500 & 4.59 & --1.51 & 0.6 & ~~1.03       & ~~0.68       &  0.35 &  0.89 & ~~0.54  \\
		HD 59392    & 6045 & 3.95 & --1.57 & 1.4 & ~~0.85       & ~~0.40       &  0.45 &  0.77 & ~~0.32  \\
		HD 97320    & 6110 & 4.26 & --1.12 & 1.4 & ~~1.11       & ~~0.75       &  0.36 &  0.58 & ~~0.22  \\
		HD 102200   & 6115 & 4.29 & --1.13 & 1.4 & ~~0.94       & ~~0.56       &  0.38 &  0.42 & ~~0.04  \\
		HD 122196   & 5985 & 3.91 & --1.68 & 1.5 & ~~0.18       & --0.28       &  0.46 &  0.21 & --0.25  \\
		HD 298986   & 6130 & 4.26 & --1.30 & 1.4 & ~~0.85       & ~~0.46       &  0.39 &  0.50 & ~~0.11  \\
		\multicolumn{10}{c}{\Teff\ from VJHK photometry \citep{dsph_parameters}} \\
		CD -24 1782 & 5140 & 2.62 & --2.67 & 1.2 & --0.54       & --1.11       &  0.57 &  0.48 & --0.09  \\
		HD 8724     & 4560 & 1.29 & --1.71 & 1.5 & ~~0.38(0.10) & --0.30(0.16) &  0.68 &  0.44 & --0.24  \\
		HD 108317   & 5270 & 2.96 & --2.13 & 1.2 & --0.01       & --0.53       &  0.52 &  0.47 & --0.05  \\
		HD 122563   & 4600 & 1.40 & --2.46 & 1.6 & --0.64       & --1.46       &  0.82 &  0.17 & --0.65  \\
		HD 128279   & 5200 & 3.00 & --2.14 & 1.1 & --0.16       & --0.66       &  0.50 &  0.33 & --0.17  \\
		HD 218857   & 5060 & 2.53 & --1.87 & 1.4 & --0.18       & --0.71       &  0.53 &  0.04 & --0.49  \\
		HE2327-5642 & 5050 & 2.20 & --2.92 & 1.7 & --0.39       & --0.96       &  0.57 &  0.88 & ~~0.31  \\
		\multicolumn{10}{c}{\Teff\ from Gaia photometry \citep{2022A&A...665A..10L}} \\
		HE1219-0312 & 5145 & 2.76 & --2.74 & 1.6 & ~~0.34       & --0.24       &  0.58 &  1.43 & ~~0.85  \\
		\hline \noalign{\smallskip}
	\end{tabular}
	
	Notes. Microturbulent velocities $\xi_t$ (\kms). The numbers in parentheses are the abundance errors $\sigma_{\rm Pd}$.
}	
\end{table*}

\section{Stellar sample, observations, atmospheric parameters} \label{Sect:sample}

\subsection{Stellar sample and observational material}

Our sample stars were selected from the stellar samples of \citet{2003A&A...397..275M,dsph_parameters} and \citet{2012A&A...545A..31H}. The main criterion was a reliable detection of, at least, one line of \ion{Pd}{i} in the star's high-resolution spectrum. We used the spectral archives of the ESO instruments: the Ultraviolet and Visual Echelle Spectrograph (UVES), the Fiber-fed Extended Range Optical Spectrograph (FEROS),
and the High Accuracy Radial velocity Planet Searcher (HARPS).
 Table~\ref{Tab:spectra} indicates the Program Ids, spectral ranges, spectral resolving power R = $\lambda/\Delta\lambda$, and the signal-to-noise ratio (S/N) corresponding to the blue spectral range, where the \ion{Pd}{i} lines are located. 
For the longer wavelengths, where lines of \ion{Fe}{ii}, \ion{Sr}{ii}, \ion{Ba}{ii}, and \ion{Eu}{ii} are located, S/N $> 100$.

The 48 stars of our sample are listed in Table~\ref{Tab:abundances1}.

\subsection{Effective temperatures}

For 13 stars, we use \Teff\ determined in our previous studies by fitting the Balmer line wings \citep[][six stars]{2003A&A...397..275M} and using the V-J, V-H, and V-K colors \citep[][seven stars]{dsph_parameters}. The uncertainty in \Teff\ was estimated as 100~K for spectroscopic determinations and from 45~K to 80~K for different stars with photometric temperatures.
For 28 stars, we adopt \Teff\ derived from the infrared flux method (IRFM) by \citet[][23 stars]{Casagrande2010,Casagrande2011}, \citet[][four stars]{2009A&A...497..497G}, and \citet[][HD~204543]{1999A&AS..139..335A}. The uncertainty in \Teff (IRFM) is, on average, 100~K. For HE1219-0312, we adopt \Teff\ = 5145$\pm$100~K as obtained by \citet{2022A&A...665A..10L} based on Gaia photometry.

For the remaining stars, we calculated photometric effective temperatures from the color-\Teff\ calibrations of \citet[][five dwarf stars]{Casagrande2010} and \citet[][a giant HD~122956]{Mucciarelli2021}.
We used the V, J, H, K magnitudes as provided by the SIMBAD\footnote{http://simbad.u-strasbg.fr/simbad/} database, the BP, RP, G magnitudes from the Gaia EDR3 catalog \citep{2021A&A...649A...1G}, the interstellar reddening maps of \citet{2021MNRAS.500.2607G} and \citet{2011ApJ...737..103S}, with the latter serving to constrain the upper limit for the color excess E$_{\rm B-V}$. The reddening estimates for other photometric bands were obtained based on the transformation relations from \citet{2014MNRAS.444..392C,2018MNRAS.479L.102C}. In these calculations,
atmospheric parameters reported by \citet{2012A&A...545A..31H} were used as an initial guess.
Using different colors yields very similar effective temperatures, such that the uncertainty in \Teff\ is estimated as 60~K for the giant star and 4~K to 75~K for the dwarfs.

\subsection{Surface gravities}

For seven stars, we adopt the spectroscopic surface gravities, \logg $_{\rm Sp}$, derived by \citet{dsph_parameters} from the requirement that the NLTE abundances from \ion{Fe}{i} and \ion{Fe}{ii} must be equal.
 The \logg $_{\rm Sp}$ values agree within 0.15~dex with the surface gravities based on the Gaia distances, \logg $_{\rm d}$, as shown by \citet{inasan_nlte}.
 The exception is HD~8724, with \logg $_{\rm Sp}$ -- \logg $_{\rm d}$ = $-0.48$, that was extensively discussed by \citet{dsph_parameters,inasan_nlte}. 
 
 For HE1219-0312, we adopted \logg\ = 2.76, as derived by \citet{2022A&A...665A..10L} using the Gaia EDR3 parallax \citep{2021A&A...649A...1G}.

For all the remaining stars, the standard relation between \logg $_{\rm d}$, \Teff, the absolute bolometric magnitude M$_{bol}$, and the stellar mass $M$ was applied. We assumed $M$ = 0.8$M_\odot$ since all our sample stars are old and still survive. The interstellar extinction was computed with $R_{\rm V} = 3.1$ and the $E(B-V)$ values either published by \citet{2012A&A...545A..31H} or obtained in this study, as described above. 
Bolometric corrections were extracted from the grids of \citet{2018MNRAS.479L.102C}. In order to compute the star's distances, we used the Gaia EDR3 parallaxes \citep{2021A&A...649A...1G}, because our sample stars are all within 1~kpc from the Sun.
For the most distant star, HD~186478 ($\pi$ = 1.0967$\pm$0.0179~mas), the distance was also calculated from the maximum of the distance probability distribution function, as recommended by \citet{2015PASP..127..994B}. 
Using this distance instead of d = 1/$\pi$ leads to a shift of 0.03~dex in the derived surface gravity. We estimate the uncertainty in \logg $_{\rm d}$ as 0.1~dex. It arises primarily from the uncertainties in \Teff\ and parallax, with the latter being insignificant for our stellar sample due to $\sigma_\pi / \pi < 1\%$. 

However, the star HD~83212 appeared the exception. With \Teff (IRFM) = 4550~K from \citet{2009A&A...497..497G} and \logg $_{\rm d}$ = 1.70, we obtained the lower NLTE abundance from lines of \ion{Fe}{i} than lines of \ion{Fe}{ii}, by 0.52~dex. 
We also checked \Teff (IRFM) = 4455~K from \citet{1999A&AS..139..335A} with the corresponding \logg $_{\rm d}$ = 1.60 and failed to remove the abundance discrepancy. Then, using \Teff\ = 4550~K and lines of \ion{Fe}{i} and \ion{Fe}{ii}, we derived \logg $_{\rm Sp}$ = 1.0. As in the case of HD~8724, it is hard to understand a source of an extremely large difference of 0.7~dex(!) between \logg $_{\rm d}$ and \logg $_{\rm Sp}$ for HD~83212. 
 The uncertainty in the Gaia EDR3 parallax $\pi$ = 1.2723$\pm$0.0211 cannot result in a large error of the distance-based surface gravity. With RUVE = 0.866, HD~83212 is unlikely a binary. 
 Further abundance analysis of HD~83212 was made with \logg $_{\rm Sp}$ = 1.0.

Hereafter, the stars with \logg\ $\le 3.0$ are referred to as giants and the stars with the greater \logg\ as dwarfs.

\begin{table}
    \centering
    \caption{\label{Tab:lines} Spectral lines used for determinations of stellar Fe, Sr, Ba, and Eu abundances.}
    \begin{tabular}{lccrc}
\hline \noalign{\smallskip}
    Species & $\lambda$, \AA & $E_{exc}$, eV & $\log gf$ & Ref. \\
\noalign{\smallskip} \hline \noalign{\smallskip}
 \ion{Fe}{ii}  & 4923.93 & 2.89 & $-$1.42 & (1) \\
 \ion{Fe}{ii}  & 5018.44 & 2.89 & $-$1.23 & (1) \\
 \ion{Fe}{ii}  & 5197.58 & 3.23 & $-2.24$ & (2) \\
 \ion{Fe}{ii}  & 5264.81 & 3.23 & $-3.02$ & (2)  \\
 \ion{Fe}{ii}  & 5284.11 & 2.89 & $-3.09$ & (2)  \\
 \ion{Fe}{ii}  & 5325.55 & 3.22 & $-3.21$ & (2)  \\
 \ion{Fe}{ii}  & 5414.07 & 3.22 & $-3.54$ & (2)  \\
 \ion{Fe}{ii}  & 5425.26 & 3.20 & $-3.28$ & (2)  \\
 \ion{Sr}{ii}  & 4077.72 & 0.00 &  0.15   & (3)  \\
 \ion{Sr}{ii}  & 4215.54 & 0.00 & $-0.17$ & (3) \\
 \ion{Ba}{ii}  & 5853.67 & 0.60 & $-1.00$ & (3) \\
 \ion{Ba}{ii}  & 6496.90 & 0.60 & $-0.38$ & (3) \\
 \ion{Eu}{ii}  & 4129.72 & 0.00 &  0.22   & (4) \\
 \ion{Eu}{ii}  & 3819.67$^1$ & 0.00 & 0.51 & (4) \\
 \ion{Eu}{ii}  & 3907.11$^1$ & 0.21 & 0.17 & (4) \\
 \ion{Eu}{ii}  & 4205.02$^2$ & 0.00 & 0.21 & (4) \\
\noalign{\smallskip} \hline \noalign{\smallskip}
\end{tabular}

References. (1) \citet{2019A&A...631A..43M}, (2) \citet{RU} corrected by $-0.11$~dex following \citet{Grevesse1999}, (3) \citet{1980wtpa.book.....R}, (4) \citet{Lawler_Eu}. 
 Notes. $^1$ only for HE2327-5642, $^2$ for HE2327-5642 and HE1219-0312.
\end{table}

\subsection{Metallicities and microturbulent velocities}\label{sect:iron}

Everywhere in this study, stellar elemental abundances were derived using the synthetic spectrum method 
with the code \textsc{synthV\_NLTE} \citep{2019ASPC} that computes the theoretical spectra and the IDL visualization program \textsc{BinMag} \citep{2018ascl.soft05015K} that makes possible to obtain the best fit to the observed spectrum. The required line list, together with atomic data, were taken from VALD.  

The iron abundances were determined from lines of \ion{Fe}{ii} (Table~\ref{Tab:lines}) under the LTE assumption. The NLTE abundance corrections, $\Delta_{\rm NLTE} = \eps{NLTE} - \eps{LTE}$, for these lines do not exceed 0.01~dex in the model atmospheres with [Fe/H] $> -3$ \citep{mash_fe}. It was obtained that our sample covers the $-2.92 \le$ [Fe/H] $\le -0.56$ metallicity range. The [Fe/H] values were computed using the Solar System abundance $\eps{Fe,met}$ = 7.45 \citep{Lodders2021}.
The dispersion in the single line measurements around the mean, $\sigma_{\rm Fe}$, amounts to 0.03 to 0.06~dex for different stars. 
The exception is HE1219-0312 with [Fe/H] = $-2.74\pm0.18$ adopted from \cite{2022A&A...665A..10L}.

Microturbulent velocities $\xi_t$ were computed using the empirical relations deduced by \citet{2015ApJ...808..148S} and \citet{dsph_parameters} for the dwarfs and the giants, respectively. 

Table~\ref{Tab:abundances1} presents the obtained atmospheric parameters. 

\section{Stellar abundances of Pd, Sr, Ba, and Eu} \label{Sect:stars}

For Pd, Sr, Ba, and Eu, we determined the NLTE and LTE abundances.
The code \textsc{synthV\_NLTE} \citep{2019ASPC} is designed to implement the departure coefficients from the code {\sc detail} and to calculate lines of the investigated species taking into account the NLTE effects, while lines of other elements under the LTE assumption.

 \begin{figure}
 \begin{center}
  \resizebox{88mm}{!}{\includegraphics{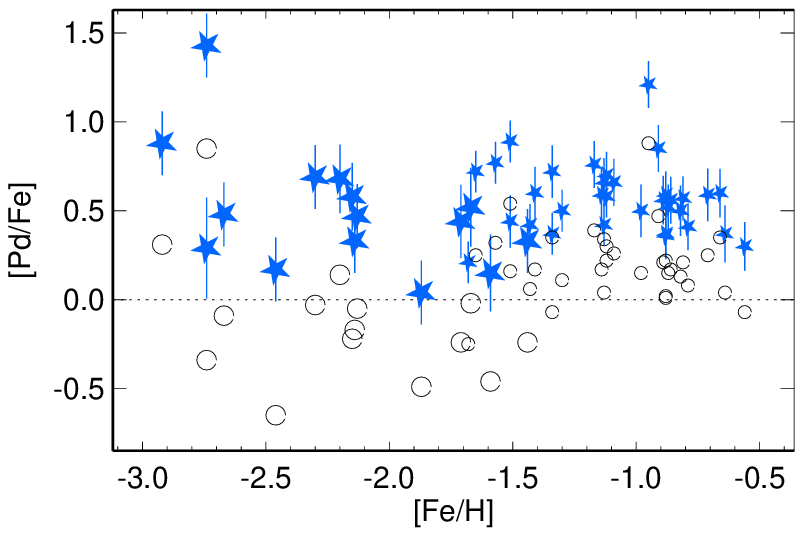}}
 \caption{NLTE (stars) and LTE (open circles) abundance ratios [Pd/Fe] for the sample stars. The smaller and bigger size symbols correspond to the dwarfs (\logg\ $> 3.0$) and the giants (\logg\ $\le 3.0$), respectively. For each star, the error bar corresponds to $\sigma_{\rm tot}$.
 }
 \label{fig:pd_nlte_lte}
 \end{center}
 \end{figure}

\subsection{Palladium}\label{sect:uncertain}

Of the five lines listed in Table~\ref{tab:sun}, \ion{Pd}{i} 3609 and 3690~\AA\ were not detected in any of our sample stars. 
In contrast, \ion{Pd}{i} 3404\AA\ was measured for each star. For 15 stars it was the only palladium line detected.
The observational errors due to the uncertainties in the continuum placement and accounting for the blending lines were evaluated as the stochastic error $\sigma_{\rm Pd}$ of the mean abundance. 
For our sample stars, $\sigma_{\rm Pd}$ ranges from 0.04~dex to 0.24~dex (Table~\ref{Tab:abundances1}).
In the case of using only \ion{Pd}{i} 3404\AA, we estimated $\sigma_{\rm Pd}$ as 0.1~dex due to the uncertainty in the continuum placement.

For each star, the NLTE calculations were performed with \kH\ = 0.1, and, for the selected stars, we checked the influence of varying \kH\ on the derived abundances (see below). The NLTE effects for \ion{Pd}{i} grow toward lower \logg\ and higher \Teff. For example, for \ion{Pd}{i} 3404\AA, $\Delta_{\rm NLTE}$ = 0.35~dex for HD~29907 (5500/4.59/$-1.51$), 0.45~dex for HD~59392 (6045/3.95/$-1.57$), and 0.61~dex for HD~204543 (4670/1.41/$-1.58$).

In line with \citet{2012A&A...545A..31H}, our sample giants have the lower LTE abundances than the dwarfs of close metallicities (Fig.~\ref{fig:pd_nlte_lte}). Here, we do not count the two  giant stars, which belong to the r-II group. Since the NLTE corrections are greater for the giants compared to the dwarfs, the abundance difference between these two group of stars was substantially removed in the NLTE calculations. In NLTE, our sample stars reveal enhancements of Pd relative to Fe, with [Pd/Fe] $\simeq$ 0.5, on average. Our two r-II stars have substantially higher enhancements, with [Pd/Fe] = 0.88 and 1.43, respectively. 

\begin{figure}
		\begin{center}
			\resizebox{88mm}{!}{\includegraphics{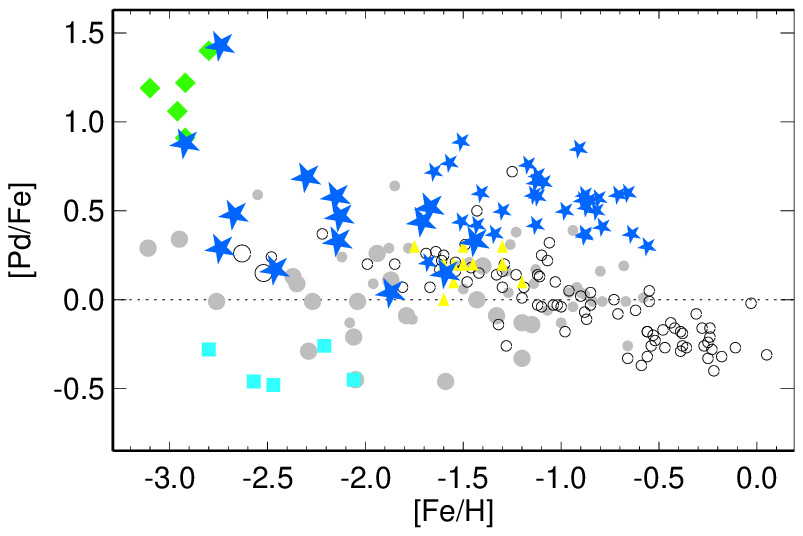}}
			\caption{NLTE abundance ratios [Pd/Fe] derived in this study (stars) compared to  the literature LTE values. Different symbols are used for different papers. Filled gray circles: \citet{2012A&A...545A..31H}, open circles: \citet{2015A&A...579A...8W}, yellow triangles: \citet{2013ApJ...768L..13P}, cyan squares: \citet{2017ApJ...837....8A}, green rhombi: five r-II stars from \citet{Sneden2003,hill2002,2017A&A...607A..91H,Lai2008,HE1219}. The smaller and bigger size symbols correspond to the dwarfs and giants, respectively.
			}
			\label{fig:pd_comparison}
		\end{center}
\end{figure}

In Fig.~\ref{fig:pd_comparison}, the obtained NLTE abundances are compared to the LTE data from \citet{2012A&A...545A..31H,2013ApJ...768L..13P, 2015A&A...579A...8W,2017ApJ...837....8A,Sneden2003,hill2002,2017A&A...607A..91H,Lai2008}, and \citet{HE1219}.
Applying the NLTE approach leads to an upward shift of the [Pd/Fe] ratios,
and this is important for better understanding the mechanisms of palladium production during the Galaxy history (see Sect.~\ref{Sect:origin}).

\subsection{Uncertainties in the Pd NLTE abundances}

\begin{table}
	\centering
	\caption{Error estimates (dex) for abundances derived from \ion{Pd}{i} 3404~\AA. }
	\label{tab:uncertainties}
	\begin{tabular}{lcc} %
		\hline\hline \noalign{\smallskip}
		& \multicolumn{1}{c}{4600/1.40/$-2.46$} & \multicolumn{1}{c}{5985/3.91/$-1.68$ } \\
		& \multicolumn{1}{c}{$\eps{Pd}$ = $-0.64$} & \multicolumn{1}{c}{$\eps{Pd}$ = 0.18} \\
		\noalign{\smallskip}\hline \noalign{\smallskip}
 & \multicolumn{2}{c}{Changes relative to $\eps{Pd}$} \\
		\multicolumn{3}{l}{Line formation treatment} \\
		LTE & $-0.82$ & $-0.46$  \\
		\kH\ = 0 & ~~0.18 & ~~0.06 \\
		\kH\ = 1 & ~~0.08 & ~~0.00 \\
		\multicolumn{3}{l}{Atmospheric parameters} \\
		$\Delta$ \Teff\ = +100~K &  ~~0.17 & ~~0.11  \\
		$\Delta$ \logg\ = +0.1   & $-0.03$ & ~~0.00 \\
\multicolumn{3}{c}{Abundance errors	} \\
	 $\sigma_{\rm Pd}$ & 0.1 & 0.1 \\
	 $\sigma_{\rm Fe}$ & ~~0.03 & ~~0.08 \\
	 $\sigma_{\rm tot}$ & ~~0.20 & ~~0.17 \\
		\noalign{\smallskip}\hline \noalign{\smallskip}
	\end{tabular}
	
		Notes. 0.00 means smaller than 0.01~dex, in absolute value. 	
\end{table}

In order to estimate the systematic abundance errors caused by applying a rough theoretical approximation for a treatment of inelastic processes in the \ion{Pd}{i} + \ion{H}{i} collisions, the NLTE calculations were performed for a number of stars using not only \kH\ = 0.1, but also \kH\ = 0 and 1. For \ion{Pd}{i} 3404\AA\ in two stars which represent dwarfs and giants, namely HD~122196 and HD~122563, Table~\ref{tab:uncertainties} indicates the abundance differences between different line formation scenarios. They are substantially larger for the giant than the dwarf model. It can be seen that applying the NLTE method is essential for determining stellar Pd abundances, independent of the adopted \kH\ magnitude. The effect of increasing hydrogenic collisional rates from \kH\ = 0.1 to \kH\ = 1 is smaller than the effect of neglecting them completely. In the case of \kH\ = 0, the Pd abundances can be higher compared to that in Table~\ref{Tab:abundances1}, by about 0.2~dex for the giant stars, while by about 0.05~dex for the dwarf stars.

We also calculated abundance errors caused by the uncertainties in \Teff\ and \logg, which are $\pm$100~K and $\pm$0.1~dex for the methods used. Variations in $\xi_t$ nearly do not affect the derived Pd abundances because the \ion{Pd}{i} lines are weak.

For plotting Fig.~\ref{fig:pd_nlte_lte}, the total uncertainty $\sigma_{\rm tot}$ in [Pd/Fe]  was computed for each star by the quadratic sum of $\sigma_{\rm Pd}$, $\sigma_{\rm Fe}$, and abundance errors due to the uncertainties in \Teff\ and \logg.
Table~\ref{tab:uncertainties} summarizes the abundance errors for two selected stars.

\begin{table*}
	\centering
\scriptsize{	
	\caption{\label{Tab:abundances2} NLTE and LTE abundance ratios [X/Fe] for Sr, Ba, and Eu.}
	\begin{tabular}{lcrrrrrr}
		\hline \noalign{\smallskip}
		\multicolumn{1}{c}{Star} & [Fe/H] & \multicolumn{2}{c}{[Sr/Fe]} & \multicolumn{2}{c}{[Ba/Fe]} & \multicolumn{2}{c}{[Eu/Fe]}  \\
		&     & \multicolumn{1}{c}{NLTE} & \multicolumn{1}{c}{LTE} & \multicolumn{1}{c}{NLTE} & \multicolumn{1}{c}{LTE} & \multicolumn{1}{c}{NLTE} & \multicolumn{1}{c}{LTE}  \\
		\noalign{\smallskip}  \hline \noalign{\smallskip}
		BD -01 2582 & --2.20 & ~~0.16(0.09) & ~~0.20(0.07) & ~~0.74(0.10) & ~~1.02(0.09) &  0.79 &  0.68  \\
		CD -33 3337 & --1.34 & ~~0.00(0.06) & ~~0.09(0.09) & --0.01(0.05) & ~~0.04(0.11) &  0.37 &  0.25  \\
		CD -57 1633 & --0.88 & --0.16(0.04) & --0.09(0.04) & --0.03(0.04) & ~~0.09(0.11) &  0.50 &  0.42  \\
		HD 3567     & --1.14 & --0.05(0.04) & ~~0.05(0.06) & ~~0.16(0.08) & ~~0.26(0.19) &  0.69 &  0.57  \\
		HD 22879    & --0.89 & ~~0.17(0.04) & ~~0.21(0.04) & ~~0.05(0.05) & ~~0.18(0.08) &  0.46 &  0.38  \\
		HD 25704    & --0.88 & --0.05(0.03) & ~~0.00(0.03) & --0.08(0.05) & ~~0.03(0.06) &  0.36 &  0.28  \\
		HD 63077    & --0.82 & ~~0.00(0.04) & ~~0.04(0.04) & --0.11(0.04) & ~~0.02(0.09) &  0.40 &  0.32  \\
		HD 63598    & --0.87 & ~~0.13(0.04) & ~~0.17(0.06) & --0.02(0.03) & ~~0.11(0.11) &  0.39 &  0.31  \\
		HD 76932    & --0.88 & ~~0.17(0.04) & ~~0.21(0.04) & ~~0.06(0.11) & ~~0.14(0.11) &  0.48 &  0.39  \\
		HD 83212    & --1.44 & ~~0.23(0.20) & ~~0.20(0.20) & ~~0.06(0.08) & ~~0.18(0.13) &  0.30 &  0.28  \\
		HD 105004   & --0.79 & --0.09(0.04) & --0.05(0.04) & --0.02(0.04) & ~~0.10(0.11) &  0.31 &  0.25  \\
		HD 106038   & --1.34 & ~~0.52(0.05) & ~~0.57(0.05) & ~~0.54(0.06) & ~~0.71(0.12) &  0.44 &  0.34  \\
		HD 111980   & --1.09 & ~~0.30(0.04) & ~~0.34(0.05) & ~~0.25(0.06) & ~~0.44(0.09) &  0.38 &  0.28  \\
		HD 113679   & --0.64 & ~~0.05(0.05) & ~~0.07(0.05) & --0.07(0.05) & ~~0.07(0.08) &  0.31 &  0.26  \\
		HD 120559   & --0.98 & --0.07(0.09) & --0.06(0.08) & --0.14(0.07) & --0.05(0.08) &  0.43 &  0.37  \\
		HD 121004   & --0.66 & ~~0.14(0.06) & ~~0.16(0.06) & ~~0.06(0.05) & ~~0.18(0.08) &  0.52 &  0.46  \\
		HD 126587   & --2.74 &              &              & --0.15(0.04) & --0.20(0.04) &       &        \\
		HD 126681   & --1.13 & ~~0.24(0.05) & ~~0.26(0.06) & ~~0.33(0.06) & ~~0.45(0.08) &  0.52 &  0.46  \\
		HD 132475   & --1.41 & ~~0.28(0.05) & ~~0.32(0.05) & ~~0.23(0.06) & ~~0.37(0.11) &  0.39 &  0.29  \\
		HD 160617   & --1.65 & --0.14(0.09) & ~~0.00(0.08) & ~~0.16(0.09) & ~~0.16(0.10) &  0.53 &  0.41  \\
		HD 165195   & --2.15 & ~~0.06(0.06) & --0.02(0.06) & --0.25(0.05) & --0.14(0.11) &  0.43 &  0.30  \\
		HD 166913   & --1.43 & ~~0.17(0.06) & ~~0.28(0.06) & ~~0.14(0.06) & ~~0.17(0.07) &       & $<$ 0.41 \\
		HD 175179   & --0.71 & ~~0.23(0.04) & ~~0.26(0.04) & ~~0.24(0.05) & ~~0.39(0.05) &  0.38 &  0.32  \\
		HD 186478   & --2.30 & --0.01(0.10) & --0.04(0.09) & --0.19(0.05) & --0.11(0.09) &  0.54 &  0.38  \\
		HD 188510   & --1.51 & --0.10(0.06) & --0.08(0.06) & ~~0.04(0.06) & ~~0.08(0.07) &  0.47 &  0.41  \\
		HD 189558   & --1.12 & ~~0.28(0.03) & ~~0.31(0.02) & ~~0.30(0.08) & ~~0.51(0.23) &  0.44 &  0.35  \\
		HD 195633   & --0.56 & ~~0.04(0.05) & ~~0.10(0.07) & --0.07(0.03) & ~~0.11(0.13) &  0.21 &  0.12  \\
		HD 204543   & --1.59 & --0.03(0.09) & --0.06(0.10) & --0.11(0.08) & ~~0.08(0.14) &  0.15 &  0.07  \\
		BD +20 0571 & --0.91 &              &              & ~~0.13(0.10) & ~~0.28(0.06) &       &        \\
		CD -45 3283 & --0.95 &              &              & ~~0.17(0.05) & ~~0.29(0.10) &       &        \\
		HD 60319    & --0.86 & ~~0.14(0.05) & ~~0.19(0.06) & --0.01(0.04) & ~~0.12(0.11) &  0.45 &  0.35  \\
		HD 103723   & --0.81 & ~~0.03(0.04) & ~~0.09(0.07) & ~~0.00(0.05) & ~~0.13(0.07) &  0.51 &  0.41  \\
		HD 122956   & --1.67 & ~~0.09(0.09) & ~~0.07(0.08) & --0.10(0.09) & ~~0.06(0.09) &  0.49 &  0.43  \\
		HD 205650   & --1.17 & ~~0.26(0.04) & ~~0.32(0.05) & ~~0.07(0.05) & ~~0.16(0.08) &  0.60 &  0.52  \\
		HD 29907    & --1.51 & --0.03(0.04) & --0.02(0.05) & --0.03(0.03) &  0.01(0.01) &  0.69 &  0.64  \\
		HD 59392    & --1.57 &  0.03(0.04)  &  0.16(0.01) &  0.16(0.02) &  0.20(0.06) &  0.67 &  0.55  \\
		HD 97320    & --1.12 & --0.12(0.04) & --0.03(0.01) & --0.03(0.00) &  0.05(0.08) &  0.36 &  0.26  \\
		HD 102200   & --1.13 & --0.28(0.06) & --0.18(0.04) & --0.08(0.02) & --0.01(0.05) &  0.53 &  0.43  \\
		HD 122196   & --1.68 & --0.51(0.01) & --0.36(0.01) & --0.17(0.03) & --0.19(0.00) &  0.11 &  0.00  \\
		HD 298986   & --1.30 & --0.27(0.00) & --0.15(0.02) & --0.07(0.03) & --0.04(0.03) &  0.54 &  0.44  \\
		CD -24 1782 & --2.67 & --0.05(0.03) & --0.01(0.00) & --0.64(0.00) & --0.68(0.00) &       &        \\
		HD 8724     & --1.71 & --0.03(0.00) & --0.07(0.00) & --0.06(0.01) &  0.12(0.07) &  0.23 &  0.17  \\
		HD 108317   & --2.13 & --0.02(0.03) &  0.05(0.05) & --0.09(0.02) & --0.06(0.05) &  0.54 &  0.44  \\
		HD 122563   & --2.46 & --0.20(0.11) & --0.24(0.07) & --1.07(0.01) & --1.11(0.01) & --0.57 & --0.72  \\
		HD 128279   & --2.14 & --0.45(0.01) & --0.38(0.03) & --0.48(0.01) & --0.49(0.04) & --0.02 & --0.12  \\
		HD 218857   & --1.87 &             &             & --0.47(0.05) & --0.44(0.05) &       &        \\
		HE2327-5642 & --2.92 &  0.08(0.01) &  0.21(0.01) &  0.40(0.09) &  0.42(0.03) &  1.19 &  1.09  \\
		HE1219-0312 & --2.74 &  0.00(0.08) &  0.09(0.06) &  0.66(0.04) &  0.74(0.04) &  1.57 &  1.47  \\
		\hline \noalign{\smallskip}
	\end{tabular} 
	
	Notes. The numbers in parentheses are the abundance errors $\sigma$.
}	
\end{table*}

 \begin{figure*}
 \begin{center}
\parbox{0.3\linewidth}{\includegraphics[scale=0.4]{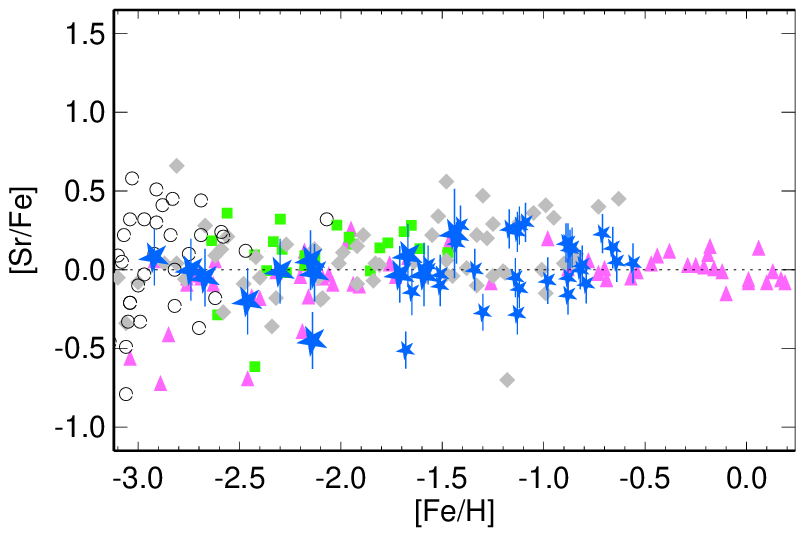}}
\parbox{0.3\linewidth}{\includegraphics[scale=0.4]{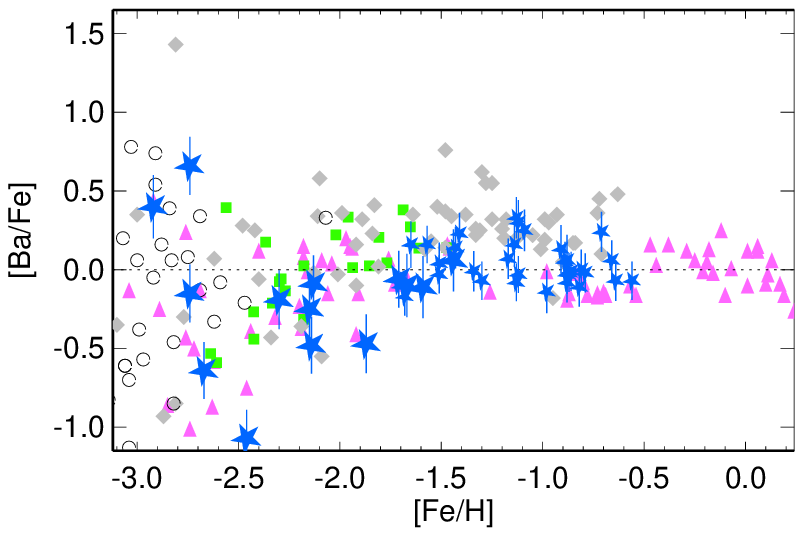}}
\parbox{0.3\linewidth}{\includegraphics[scale=0.4]{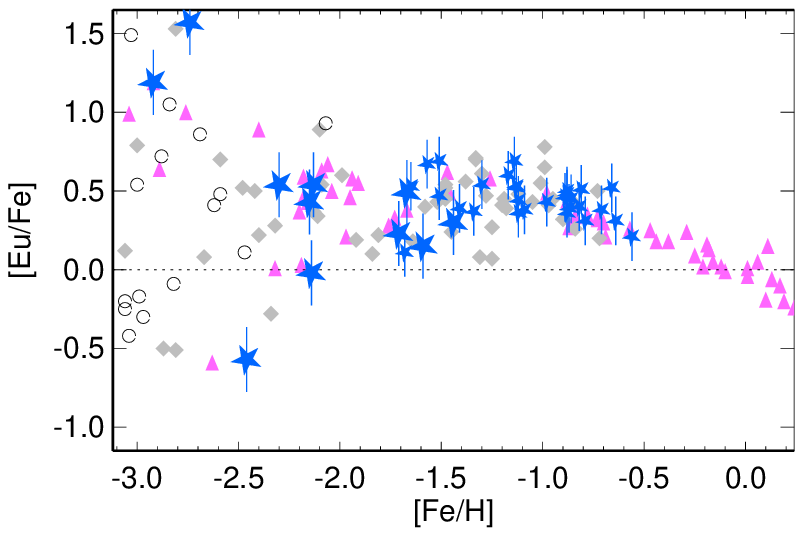}}
 \caption{Abundance ratios of Sr, Ba, Eu to Fe obtained in this study (NLTE, stars) compared to the corresponding ratios of the comparison stellar samples: magenda triangles, \citet{lick_paperII} + \citet{2017A&A...608A..89M}; gray diamonds, \citet{2012A&A...545A..31H}; green squares, \citet{2017A&A...608A..46R}; open circles, \citet{Francois2007} + \citet{2011A&A...530A.105A}.  }
 \label{fig:sr_ba_eu}
 \end{center}
 \end{figure*}

 \begin{figure*}
 \begin{center}
\resizebox{140mm}{!}{\includegraphics{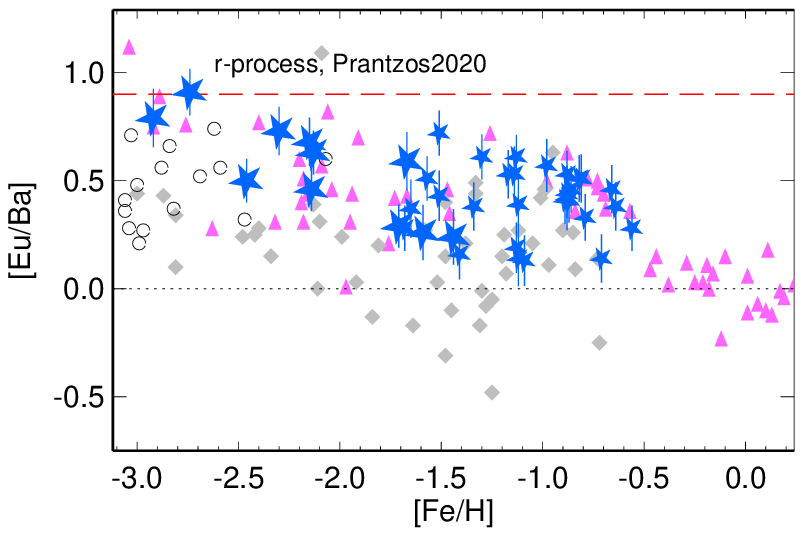} \includegraphics{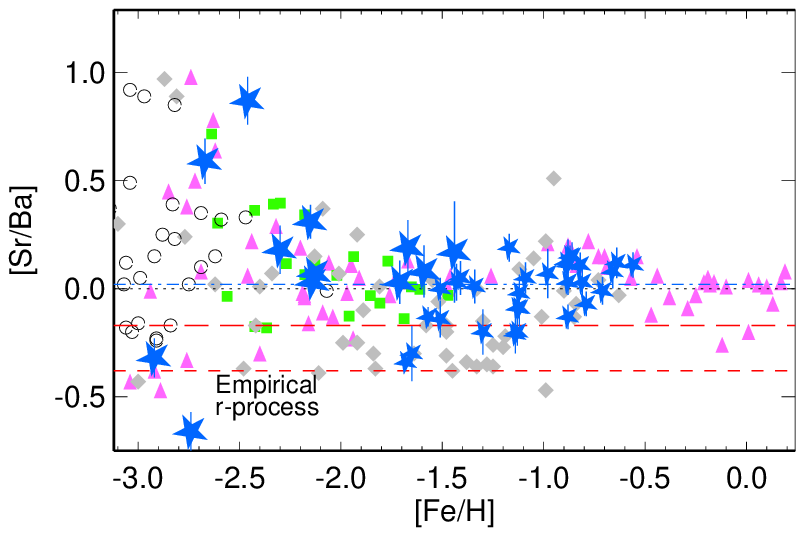}} 
\resizebox{140mm}{!}{\includegraphics{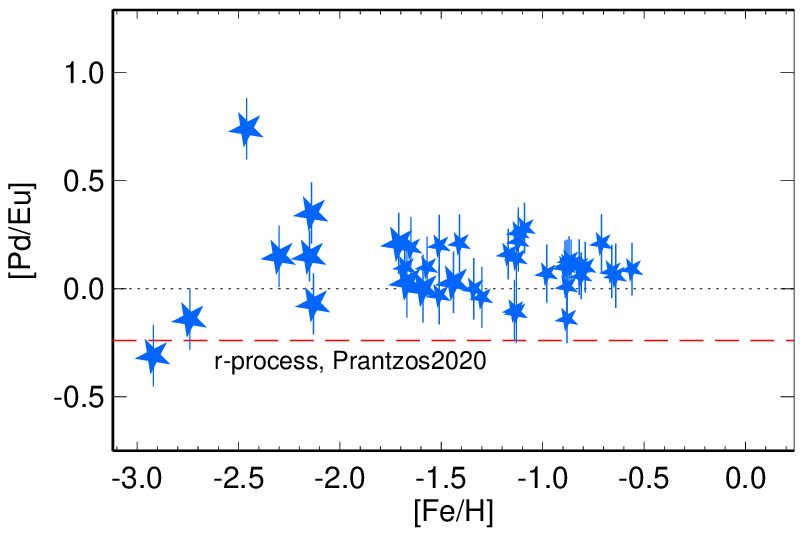} \includegraphics{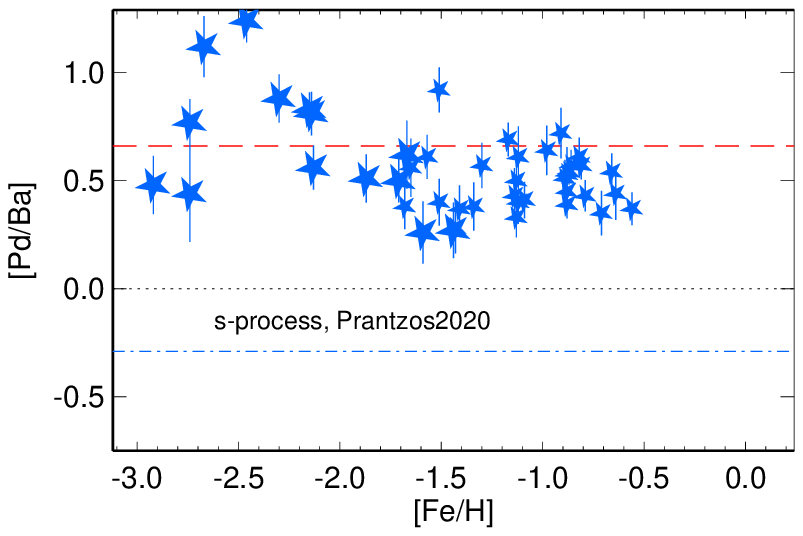}}
 \caption{Same as in Fig.~\ref{fig:sr_ba_eu} for the abundance ratios among Eu, Ba, Sr, and Pd. The long-dashed red and dash-dotted blue lines indicate the elemental ratios predicted by \citet{Prantzos2020} for pure r- and s-process, respectively. The short-dashed red line corresponds to the empirical r-process ratio [Sr/Ba]$_{\rm r-II} = -0.38$. }
 \label{fig:ratios}
 \end{center}
 \end{figure*}


\subsection{Strontium, barium, europium}

Spectral lines of \ion{Sr}{ii}, \ion{Ba}{ii}, and \ion{Eu}{ii} used in abundance analyses are listed in Table~\ref{Tab:lines}.
 Each of these lines is composed of the isotopic and HFS components. Their wavelengths and $gf$-values along with the van der Waals damping constants were taken from a recent version of VALD \citep{2019ARep...63.1010P}. We employed the Solar System isotopic fractions from \citet{Lodders2021}. The derived abundances depend only weakly on the isotopic mixture adopted in the calculations. 

For the four stars, we do not have spectra covering lines of \ion{Eu}{ii} and \ion{Sr}{ii}. Lines of \ion{Eu}{ii} cannot be extracted from noise in the spectrum of CD~-24~1782 with [Fe/H] = $-2.67$. Spectrum of HD~166913 around the \ion{Eu}{ii} 4129 and 4205~\AA\ lines is affected by absorptions of unknown origin, such that even the upper limit of the Eu abundance was determined uncertainly.

The NLTE calculations were performed using the methods treated in our previous studies: \citet[][\ion{Eu}{ii}]{mash_eu},  \citet[][\ion{Ba}{ii}]{2019AstL...45..341M}, and 
 \citet[][\ion{Sr}{ii}]{2022MNRAS.509.3626M}.
The obtained LTE and NLTE [X/Fe] ratios are presented in Table~\ref{Tab:abundances2}. The Solar System abundances were taken from \citet{Lodders2021}: $\eps{Sr,met}$ = 2.88, $\eps{Ba,met}$ = 2.17, $\eps{Eu,met}$ = 0.51. 

\subsection{Notes on individual stars}

Three stars of our sample are not shown in Figs.~\ref{fig:sr_ba_eu} and \ref{fig:ratios}, and they are excluded from further discussion because their chemical abundances indicate nucleosynthetic origins that are distinct from those of the chemical elements in the halo stars of similar metallicity. The arguments are as follows.

(i) The star BD-01~2582 was identified as a carbon enhanced (CEMP) star by \citet{1982PASP...94...55K}. 
Based on the abundances [C/Fe] = 1.02, [Ba/Fe] = 1.05, [Eu/Fe] = 0.36 determined by \citet{roe14}, 
 \citet{yoon16} classified the star as CEMP-s. 
\citet{2022ApJ...927...13Z}  obtained very similar C enhancement 
([C/Fe] = 1.03), but a high enhancement of Eu relative to Ba ([Eu/Ba] = 0.57) and classified the star as CEMP-r. In this study, we confirmed that BD-01~2582 is a CEMP star, with [C/Fe] = 0.97, but its NLTE abundances of Ba and Eu support neither CEMP-s, nor CEMP-r classification. With similar enhancements of Ba and Eu relative to Fe, BD-01 2582 can be classified as CEMP-r/s. Detailed abundance analysis of this star will be presented in a separate paper. 

(ii) The star HD~106038 is known as a Be-rich halo dwarf \citep{2008MNRAS.385L..93S}. Such a phenomenon can be explained if the star was formed in the vicinity of a hypernova. However, \citet{2008MNRAS.385L..93S} note that HD~106038 is also enhanced with Ba and this may pose a problem for the hypernova scenario. In this study, HD~106038 was found to be enhanced with Sr, Ba, and Eu at a similar level of 0.5~dex. Such a heavy element abundance pattern is distinct from that for other stars of similar metallicity ([Fe/H] = $-1.34$) and is unlikely explained within the collapsar/hypernova scenario \citep[see, for example,][]{2024ApJ...962...68A}. An origin of chemical elements in HD~106038 remains an open question.

(iii) The star CD-45~3283 can be of extragalactic origin. Its [O/Fe] = $-0.5$ value determined by \citet{2014A&A...568A..25N} is one order of magnitude lower compared to that for the halo stars of close metallicity ([Fe/H] = $-0.95$), while similarly low [$\alpha$/Fe] ratios, including [O/Fe], are found at the same metallicities in the dwarf spheroidal (dSph) galaxies \citep[e.g. Sculptor dSph,][]{Hill_Scl}. The Ba abundance obtained for CD-45~3283 in this study, [Ba/Fe] = 0.17, is not exceptional for mildly MP stars in both our Galaxy and the Galaxy satellites. Abundances of Sr and Eu were not determined due to an absence of observed spectra.

\section{Galactic trends of the elemental ratios}\label{Sect:origin}

In this section, we first investigate origins of Eu, Ba, and Sr in our sample stars and then discuss an origin of Pd 
based on analysis of the abundance ratios involving Pd. For comparison with observations, we apply the s-process calculations of \citet{Prantzos2020} that include contributions from intermediate mass stars (main s-process) and massive rotating stars (weak s-process). Using the Solar System abundances, \citet{Prantzos2020} provide fractions of individual isotopes and elements, which  originated from the s-process, and the r-residuals obtained by subtraction of the s-process contributions from the total abundances.
 Based on the predicted r-residuals, we computed the r-process elemental ratios: [Eu/Ba]$_{\rm r}$ = 0.90, [Sr/Ba]$_{\rm r} = -0.17$, [Pd/Eu]$_{\rm r} = -0.24$, and [Pd/Ba]$_{\rm r}$ = 0.66. The r-II stars serve as a benchmark for the abundance pattern of heavy elements produced by the pure r-process. Elemental ratios obtained by averaging the data for the best representatives of the r-II stars: [Eu/Ba]$_{\rm r-II}$ = 0.88 \citep[][eight stars]{HERESIX} and [Sr/Ba]$_{\rm r-II} = -0.38$ \citep[][six stars]{2017A&A...608A..89M} are referred here to as the empirical r-process ratios.
 
 In the [Fe/H] $> -2$ range, our stellar sample reveals the trends for the elemental ratios involving Eu, Ba, and Sr (Figs.~\ref{fig:sr_ba_eu} and \ref{fig:ratios}), 
   which are typical for the Galactic halo and thick disk stellar populations. The comparison stellar samples were taken from \citet{Francois2007,2011A&A...530A.105A,2012A&A...545A..31H,lick_paperII,2017A&A...608A..89M}, and \citet{2017A&A...608A..46R}.
   Europium is enhanced relative to Fe, and there is a clear decline of [Eu/Fe] for [Fe/H] $> -1$ that indicates the onset of Fe production by type Ia supernovae (SNeIa). Europium is enhanced relative to Ba, and [Eu/Ba] decreases toward higher metallicity that means increasing contribution of the main s-process in AGB stars to Ba production. The [Sr/Fe], [Ba/Fe], and [Sr/Ba] ratios are close to the solar ones, indicating a common origin of Sr and Ba in the r- and main s-process. 
 
 In the [Fe/H] $> -2$ range, Pd is enhanced relative to Fe and Ba and tightly correlates with Eu (Fig.~\ref{fig:ratios}). By comparing the observed average [Pd/Eu] = 0.09$\pm$0.10 with [Pd/Eu]$_{\rm r} = -0.24$ and [Pd/Eu]$_{\rm s}$ = 0.97 \citep{Prantzos2020}, we estimated  contributions of the r- and s-process to the Pd abundances of our $-1.71 \le$ [Fe/H] $\le -0.56$ stars as approximately 70\%\ and 30\%.
 
In the very metal-poor (VMP, [Fe/H] $< -2$) region, Fig.~\ref{fig:sr_ba_eu} shows a large spread of [Eu/Fe] and [Ba/Fe], 
suggesting the rarity of the nucleosynthesis events for heavy elements and incomplete mixing of produced nuclei.
 Despite our two r-II stars ([Eu/Fe] $>$ 1), the star HD~122563 ([Eu/Fe] = $-0.57$) known as r-process poor, and HD~128279 ([Eu/Fe] = $-0.02$) have very different [Eu/Fe] ratios,
they align well with the remaining sample stars on the [Eu/Ba] versus [Fe/H] plane (Fig.~\ref{fig:ratios}) and reveal dominant (the r-II stars) or significant contribution of the r-process to their heavy element abundances. 

The [Pd/Eu] and [Pd/Ba] ratios of our three most MP stars (HE2327-5642, HE1219-0312, HD~126587) provide the clear evidence for a pure r-process origin of Pd in these stars. Dominant contribution of the r-process to the Pd abundance is revealed by HD~108317 ([Fe/H] = $-2.13$, [Pd/Eu] = $-0.07\pm$0.14, [Pd/Ba] =0.56$\pm$0.10). 

	However, palladium is strongly enhanced relative to Ba and Eu in HD~122563 and CD~-24~1782 (only Ba was measured). Exactly these two stars have the high [Sr/Ba] $>$ 0.5 ratios suggesting the contribution to their Sr abundances from an extra source besides the r-process. Based on determinations of the odd-to-even isotope abundance ratio for barium, \citet{2025A&A...699A.262S} provide the evidence for the early (or non-standard) s-process in extremely metal-poor (EMP, [Fe/H] $< -3$), fast rotating massive stars \citep{2008ApJ...687L..95P,2014A&A...565A..51C} being the source of extra Sr in HD~122563 and estimate the contributions of the r-process and the early s-process as 55\%\ and 45\%, respectively. Calculations of the non-standard s-process in the [Fe/H] = $-1.8$ models of 20 to 60~$M_\odot$ show that the production factors for Pd grow by 2-3~dex in the rotating (40\%\ of the critical velocity) models compared to the non-rotating ones \citep{2018A&A...618A.133C}. The increase is greater, more than 4~dex, in the [Fe/H] = $-3.8$ models \citep{2016MNRAS.456.1803F}.
We propose that nucleosynthesis in fast rotating massive stars contributed to the abundance of not only Sr, but also Pd in HD~122563. It is natural to assume that the matter out of which the star CD~-24~1782 formed was enriched by similar sources. 

\section{Conclusions}\label{Sect:conclusion}

We constructed a new, comprehensive model atom for \ion{Pd}{i} and, for the first time, derived the NLTE abundances of Pd in the Sun and the sample of 48 stars covering the $-2.92 \le$ [Fe/H] $\le -0.56$ metallicity range. 
In the NLTE calculations, the \ion{Pd}{i} + \ion{H}{i} collisions were treated by applying the formulas of \citet{Steenbock1984} with the scaling factor \kH. NLTE leads to weakened lines of \ion{Pd}{i} and positive NLTE abundance corrections which grow toward higher \Teff\ and lower \logg\ up to $\Delta_{\rm NLTE}$(\kH\ = 0.1) = 0.47~dex for the 6050/3.90/$-1.65$ model and $\Delta_{\rm NLTE}$(\kH\ = 0.1) = 0.82~dex for the 4600/1.40/$-2.46$ model. In the case of pure electron collisions, $\Delta_{\rm NLTE}$ increases by 0.05~dex and 0.2~dex for the \logg\ $\succsim 4$ and \logg\ $< 2$ models, respectively.

The solar Pd abundance was derived from the \ion{Pd}{i} 3242, 3404, and 3609~\AA\ lines in the solar disc-center intensity spectrum \citep{Delbouille1973} using the classical MARCS model atmosphere \citep{Gustafssonetal:2008}. The obtained NLTE abundances $\eps{Pd,\odot}$(\kH\ = 0.1) = 1.70$\pm$0.02 and $\eps{Pd,\odot}$(\kH\ = 1) = 1.61$\pm$0.02 are consistent within the error bars with the meteoritic value $\eps{Pd,met}$ = 1.65$\pm$0.02 \citep{Lodders2021}, while the LTE abundance is lower by 0.16~dex.

Using the high-resolution and high S/N spectra from the ESO archives, we selected 48 stars with measurable lines of \ion{Pd}{i}. 
In determinations of stellar atmospheric parameters, we relied on the best available methods and data. The effective temperatures were taken from our previous studies 
and determinations of \Teff (IRFM) by \citet{Casagrande2010,Casagrande2011,2009A&A...497..497G}, and \citet{1999A&AS..139..335A}. For the six stars with missing \Teff (IRFM), we employed the color-\Teff\ calibrations of \citet{Casagrande2010} and \citet{Mucciarelli2021}. The surface gravities are based on the Gaia EDR3 parallaxes \citep{2021A&A...649A...1G}. The exceptions are seven stars from \citet{dsph_parameters} and HD~83212 with the spectroscopic surface gravities. 
The iron abundances were determined from the \ion{Fe}{ii} lines.

The NLTE abundances were determined for Pd and also Eu, Ba and Sr that are the best representatives of the elements produced effectively in the r-process, the main s-process, and the weak s-process, respectively. Our findings can be summarized as follows.
\begin{itemize}
	\item For Pd, NLTE largely removes the discrepancies in the LTE abundances between the giant and dwarf stars of similar metallicities.
	\item Palladium is enhanced relative to Fe over the whole metallicity range investigated.
	\item Palladium tightly correlates with Eu in the $-1.71 \le$ [Fe/H] $\le -0.56$ stars. Contributions of the r- and s-processes to their Pd abundances were estimated as 70\%\ and 30\%, using the predictions of \citet{Prantzos2020} for the solar r/s ratio.
	\item Palladium is of pure r-process origin in our two r-II stars. Two another stars in the [Fe/H] $< -2$ range reveal dominant contribution of the r-process to their Pd abundances.
	\item The stars HD~122563 and CD~-24~1782 are strongly enhanced with not only Sr, but also Pd relative to Ba and Eu, suggesting the contributions to Sr and Pd from a distinct channel, besides the r-process. Based on results of \citet{2025A&A...699A.262S}, 
	we propose that the source of extra Pd in these two stars are VMP, fast rotating massive stars.  
\end{itemize}

We conclude that using the NLTE approach is essential for obtaining accurate stellar abundances that can serve the observational constraints to the models of the Galactic Pd evolution.

\begin{acknowledgements}
We made use of the ESO UVES, HARPS, and FEROS spectral archives, the MARCS, NIST, SIMBAD, VALD, and ADS\footnote{http://adsabs.harvard.edu/abstract\_service.html} databases.
\end{acknowledgements}

\bibliographystyle{aa}
\bibliography{aa57263-25}

\clearpage

\begin{appendix}
\onecolumn
\section{Observational material}

\begin{table*}  
	\centering
\caption{\label{Tab:spectra} Sources and characteristics of the used observational material taken from the ESO archives.} 
	\begin{tabular}{rllccc}
\hline\hline \noalign{\smallskip}
N & Instrument & Program Id & Spectral range (\AA) & R & S/N$^a$ \\
\noalign{\smallskip} \hline \noalign{\smallskip}
1.  &  UVES    & UVESPOP      & 3040 -- 10400 & 88\,000   &  460 \\
2.  &  UVES    & 65.L-0507(A) & 3040 -- 6835  & 41\,000, 51\,000 & $>$ 150 \\
3. &  UVES    & 67.D-0086(A) & 3280 -- 6690 & 50\,000 & 100  \\
4.  &  UVES    & 67.D-0439(A) & 3024 -- 6835 & 50\,000, 57\,000 & $>$ 250  \\ 
5.  &  UVES    & 68.D-0094(A) & 3040 -- 6835  & 41\,000, 51\,000 & $>$ 200 \\
6. &  UVES    & 68.B-0475(A) & 3040 -- 6835  & 37\,000, 46\,000 & $>$ 100 \\
7. &  UVES    & 68.D-0546(A) & 3030 -- 6830 & 41\,000, 57\,000 & 100  \\
8. &  UVES    & 71.B-0529(A) & 3043 -- 6808 & 41\,000, 45\,000 & 100 \\
9. &  UVES    & 072.B-0585(A) & 3030 -- 6835 & 41\,000, 45\,000 & $>$ 100 \\
10.&  UVES    & 076.B-0055(A) & 3732 -- 5000  &  41\,000 & 290 \\
11. &  UVES    & 165.N-0276(A) & 3084 -- 6835 & 68\,000, 81\,000 &  80 \\
12. &  UVES    & 165.L-0263(A) & 3377 -- 4653 & 41\,000 & 120 \\
13. &  UVES    & 165.L-0263(B) & 3559 -- 4845  & 41\,000 & $>$ 150 \\
14. &  UVES    & 170.D-0010   & 3030 -- 6840 & 71\,000   & 24 \\ 
15. &  UVES    & 188.B-3002(C) & 4180 -- 6213 & 47\,000  & $>$ 200 \\
16. &  UVES    & 266.D-5655(A) & 3024 -- 6835 & 65\,000, 74\,000   & 90 \\
17. &  UVES    & 280.D-5011   & 3330 -- 6800 & 60\,000            & 20  \\ 
18. &  UVES    & 109.22VP.001 & 3282 -- 6686 & 58\,600, 66\,300   & $>$ 100 \\
19. &  UVES    & 109.22VP.002 & 3282 -- 4563 & 58\,600   & 117 \\
20. &  UVES    & 110.240W.001 & 3282 -- 6686 & 58\,600, 66\,300   & $>$ 160 \\
21. &  UVES    & 110.240W.002  & 3282 -- 4563 & 58\,600 & 145 \\
22  & HARPS   & 60.A-9700(G)  & 3780 -- 6910 & 115\,000          & 175 \\
23  & HARPS   & 072.C-0488(E) & 3780 -- 6910 & 115\,000          & $>$ 75 \\
24  & HARPS   & 096.D-0402(A) & 3780 -- 6910 & 115\,000          & 84 \\
25  & FEROS   & 072.D-0315(A) & 3528 -- 9218 & 48\,000          & 140 \\
\noalign{\smallskip}\hline \noalign{\smallskip}
\end{tabular} 

Notes. $^a$ The signal-to-noise ratio is given for $\lambda \sim$ 3400~\AA. For the longer wavelengths, up to 6500~\AA, S/N $> 100$. UVESPOP = \citet{2003Msngr.114...10B}.
\end{table*}
\twocolumn
\end{appendix}

\end{document}